\documentclass[aps,preprint,amssymb,12pt,floatfix]{revtex4}
\usepackage{subfigure}
\usepackage{lscape,graphicx}
\usepackage{rotating}

\begin{document}
\title{Allosteric communication in Dihydrofolate Reductase: Signaling network and pathways for closed to occluded transition and back}
\author{Jie Chen$^1$, Ruxandra I. Dima$^3$ and D. Thirumalai$^{1,2}$}
\thanks{Corresponding author phone: 301-405-4803; fax: 301-314-9404; thirum@glue.umd.edu}
\address{$^1$Biophysics Program,Institute for Physical Science and Technology\\
$^2$Department of Chemistry and Biochemistry, University of Maryland, College Park, MD 20742\\
$^3$Department of Chemistry, University of Cincinnati, Cincinnati, OH 45221\\}

\date{\small \today}
 
\baselineskip = 22pt

\begin{abstract}
\textit{E. Coli.} dihydrofolate reductase (DHFR) catalyzes the reduction of dihydrofolate to tetrahydrofolate.  During the catalytic cycle, DHFR undergoes conformational transitions between the closed (CS) and occluded (OS) states which, respectively, describe whether the active site is closed or occluded by the Met20 loop. The CS$\rightarrow$OS and the reverse transition may be viewed as allosteric transitions. Using a sequence-based approach we identify a network of residues that represents the allostery wiring diagram. Many of the residues in the allostery wiring diagram, that are dispersed throughout the adenosine binding domain as well as the loop domain, are not conserved. Several of the residues in the network have been previously shown by NMR experiments, mutational studies, and molecular dynamics simulations to be linked to equilibration conformational fluctuations of DHFR. To further probe the nature of events that occur during conformational fluctuations we use a self-organized polymer model to monitor the kinetics of  the CS$\rightarrow$OS  and the reverse transitions.  During the CS$\rightarrow$OS transition, coordinated changes in a number of residues in the loop domain enable the Met20 loop to slide along the $\alpha$-helix in the adenosine binding domain. Sliding is triggered by pulling of the Met20 loop by the $\beta$G-$\beta$H loop and pushing action of the $\beta$G-$\beta$H loop. The residues that facilitate the Met20 loop motion are part of the network of residues that transmit allosteric signals during the CS$\rightarrow$OS transition. Replacement of M16 and G121, whose $C_\alpha$ atoms are about 4.3$\AA$ in the CS, by a disulfide crosslink impedes that CS$\rightarrow$OS transitions. The order of events in the OS$\rightarrow$CS transition is not the reverse of the forward transition. The contact Glu18-Ser49 in the OS state persists until the sliding of the Met20 loop is nearly completed. The ensemble of structures in the transition state (TS) in both the allosteric transitions are heterogeneous. The most probable TS structure resembles the OS (CS) in the CS$\rightarrow$OS (OS$\rightarrow$CS) transition which is in accord with the Hammond postulate. Structures resembling the OS (CS) are present as minor ($\sim (1-3) \%$) component in equilibrated CS (OS) structures.  
\end{abstract}

\maketitle

\section*{Introduction}

Conformational fluctuations of proteins have been argued to play a central role in 
enzyme catalysis \cite{Schnell_Annu_Rev_04,Boehr_Science_06,Hammes_Nature_64,Schiffer_Benkovic_Annu_Rev_06}. 
Such a concept is appealing because the energy landscape of enzymes even in the folded state is rugged \cite{Hyeon_Biochem_05}, and hence thermal energy might be sufficient to access several conformational substates during a typical reaction cycle \cite{Boehr_Science_06}. In recent years, results from a number of studies have been used to propose that dynamic motions in a network of residues that promote catalytically-relevant structural transitions may be encoded in the protein structure \cite{Schiffer_PNAS_02,Schiffer_JPCB_02,Schiffer_JPCB_04,Schiffer_PNAS_05,Benkovic_Biochem_02,Cannon_NSB_96,Brooks_JACS_00,Brooks_PNAS_03,Brooks_JPC_03,Brooks_Proteins_04,Gao_Biochem_03}. While it is difficult to unambiguously demonstrate whether collective dynamics involving a network of residues facilitates catalysis \cite{Warshel_ChemRev_06}, it is clear that enzymes sample a number of distinct states during a reaction cycle. In the best studied example of \textit{E. coli} dihydrofolate reductase (DHFR) the role of the conformational motions in the enzyme in facilitating  the hydride transfer has been linked using mutational studies \cite{Benkovic_Biochem_97,Benkovic_Biochem_02,Benkovic_PNAS_06}, NMR relaxation dispersion measurements \cite{Feeney_Biochem_99,Wright_Biochem_01,Wright_PNAS_05} that probe the dynamics on $\mu s$ to $ms$ time scale, molecular dynamics simulations \cite{Brooks_PNAS_03,Schiffer_PNAS_02,Schiffer_JPCB_02}, and sequence analysis \cite{Schiffer_PNAS_02}. 

The emphasis on correlated motions on enzyme catalysis has been repeatedly questioned by Warshel and coworkers \cite{Warshel_ChemRev_06} who have shown that catalytic rates are largely affected by changes in the free energy barriers ($\Delta g^\ddagger$) in the chemical reaction step \cite{Warshel_JPC_91,Warshel_QRB_01,Warshel_JPC_01,Warshel_Biochem_07}. Thus, mutational effects or other constraints simply alter $\Delta g^\ddagger$ and hence the catalytic rates. Complex enzyme motion in a multidimensional free energy landscape is to a large extent orthogonal to the dynamics along the optimized reaction coordinate. Regardless of the rate of correlated dynamical motions on the rates of hydride transfer in DHFR, it is known that the enzyme cycles through a number of states during the catalytic cycle. The dynamics of transition between such states (referred to as allosteric states) is the topic of interest in this study. Whether the time scales in such conformational transitions occur are linked to catalysis is unclear \cite{Warshel_ChemRev_06}.

DHFR catalyzes the reduction of 7,8 dihydrofolate (DHF) to 5,6,7,8 tetrahydrofolate (THF) \cite{Schnell_Annu_Rev_04}. By binding the co-factor, nicotinamide adenine dinucleotide phosphate (NADPH), hydride transfer 
from NADPH to protonated DHF leads to production of NADP$^+$ and THF \cite{Benkovic_ChemBiol_98}. DHFR, which is required for normal folate metabolism in prokaryotes and eukaryotes, plays an important role in cell growth and proliferation in prokaryotes and eukaryotes \cite{Berg_Biochem}. As the result of its obvious clinical importance, it has been studied extensively using a wide range of experimental and theoretical methods \cite{Schnell_Annu_Rev_04}.

High resolution crystal structures show  that the \textit{E. Coli} DHFR enzyme has  eight $\beta$-strands 
 and four ${\alpha}$-helices interspersed with flexible loops that connect the secondary structural elements \cite{Matthews_Science_77,Kraut_Biochem_97}. The structure of DHFR can be partitioned into adenosine binding and loop subdomains \cite{Kraut_Biochem_97}. In the catalytic cycle, 
Met20 loop changes conformation between closed (CS) and occluded (OS) states (Fig. \ref{fig:catalysis_cycle}). Interactions through  hydrogen bond network with the $\beta$F-$\beta$G 
loop (residues 117-131) stabilize the CS \cite{Benkovic_Wright_Biochem_04}.

The crystal structures of E.{\it coli} DHFR complexes 
in the catalytic cycle have given a detailed map of the structural changes that occur in the enzyme \cite {Kraut_Biochem_97}. In addition, the conformational changes in E.{\it coli} DHFR in response to binding have been inferred  using various experimental techniques, including X-ray crystallography, fluorescence, nuclear magnetic resonance (NMR) \cite{Matthews_Science_77,Kraut_Biochem_97,Wright_Biochem_01}. Comparison of the CS and OS structures shows that the conformations of Met20 loop undergoes the largest  change during the reaction cycle. As a result, the states of DHFR  are classified using the conformations of the  Met20 loop. The active site is either closed, or occluded depending on the conformation of the Met20 loop (Fig. \ref{fig:catalysis_cycle}).  Thus, the motion of the
Met20 loop coordinates the dynamical changes in DHFR during the different stages of the catalytic cycle.

Although the structures of the CS and OS states are known the dynamic pathways connecting the two allosteric states have not been  characterized \cite{Schnell_Annu_Rev_04,Boehr_Science_06}. In this paper, we address the following questions: (1) Can the evolutionary footprints in the DHFR family of sequences be used to obtain a network of residues in DHFR that is linked to the CS$\rightarrow$OS and the reverse transition in the enzyme?  If so, what role do these residues play in the kinetics of CS$\rightarrow$OS and the reverse transitions? (b) What are the pathways and the nature of the kinetics associated with transition from OS to CS and back? (c)  What are the structures of the transition state ensemble in the OS$\rightarrow$CS transition and in the reverse reaction? 
We use a combination of bioinformatics  methods \cite{RR_Science_99,Dima_ProtSci_06},  and Brownian dynamics simulations of coarse-grained models of DHFR to address these questions. It should be emphasized that our study focuses only on the kinetics of CS$\rightarrow$OS and OS$\rightarrow$CS transitions, and not on whether the motions that drive these transitions affect hydride transfer reactions. The precise linkage between equilibrium or dynamics motions of proteins and catalysis continues to be a topic of debate \cite{Schiffer_Benkovic_Annu_Rev_06,Warshel_ChemRev_06}.

In order to determine the network of residues in DHFR that regulates the allosteric transitions we adopt a sequence-based method \cite{Dima_ProtSci_06}, which is based on the Statistical Coupling Analysis (SCA) \cite{RR_NSB_03,RR_PNAS_03,RR_PNAS_04,RR_cell_04}.
The SCA identifies many residues that are dispersed between the two subdomains as being relevant in the function of DHFR. Although several of these residues are not strongly conserved, they are predicted to covary across the DHFR family. 
In order to probe allostery in DHFR, we  carried out simulations using coarse-grained Self-organized polymer (SOP) model \cite{Hyeon_BiophyJ_07}.  The Brownian dynamics simulations reveal the dynamical changes that occur during the CS$\rightarrow$OS and OS$\rightarrow$CS transitions. The conformational changes in the Met20 loop, which occur by a  sliding motion along a helix in the adenosine binding domain, is preceded by coordinated rupture of interactions between Met20 and $\beta$F-$\beta$G loops and the formation of contacts between Met20 and $\beta$G-$\beta$H loops. Simulations in which Met16 and Gly121 are crosslinked by a disulfide bond, show that the CS$\rightarrow$OS transition is dramatically affected. In accord with the recent NMR experiments \cite{Boehr_Science_06,Clore_Nature_06}, we find a small ($\sim (1-3)\%$) of OS (CS) structures are populated by thermal fluctuations when DHFR is in the CS (OS) state. The structures of the transition state ensemble (TSE) is broad both in the forward and reverse direction. The presence of broad TSE and small barrier separating  the CS and OS states supports the conformational selection model that posits that due to the heterogeneous nature of fluctuations conformations resembling the OS state are present in the CS and vice versa.

{\bf Results and Discussion}

{\bf Allostery wiring diagram shows that key residues are dispersed throughout the structure}

We obtained 526 sequences for the DHFR family  from Pfam \cite{Pfam} (entry 00186), and 
realigned them using the  Clustalw package \cite{ClustalW}. 
We manually deleted certain sequences, and generated a multiple sequence alignment (MSA) that contained  462 sequences. 
Each of the 462 sequences has 323 residues including gaps. With the fraction of the sequences in the subalignment set to $f$=0.35 (see Methods) in the SCA, there are 74 allowed perturbations ($S_j = 0$ for $j = 1, 2, 3 ...74$)  at the 
various positions in the DHFR family.  We used the clustering protocol \cite{Blatt_PRL_96,Domany_PhysicaA_99} to identify the set of co-varying residues. After rescaling the 
$\Delta \Delta G_{ij}$ matrix (Eq. \ref{eq:delta_Gi} in Methods) ($i = 1, 2, 3, ...158$ and $j = 1, 2, 3,....74$), and using the Euclidean similarity 
measure in the coupled two-way clustering algorithm \cite{Dima_ProtSci_06} we obtained a cluster of 21 residues and a cluster 
of 19 perturbations.  As in our previous work \cite{Dima_ProtSci_06}, we propose 
that the residues that are clustered both in positions and 
perturbations constitute the minimal robust network of residues that signal the kinetics of the CS$\rightarrow$ 
OS transition and back. 
The relevant network of spatially separated residues constitutes an allostery wiring diagram (Fig. \ref{fig:conservation}A), and may encode for the promoting motions. 

To determine if the residues in the network predicted by SCA merely reflect sequence conservation we calculated the sequence 
entropy $S_i=-\sum_{x=1}^{20}{p_i^xln(p_i^x)}$.  For a perfectly conserved residue,  $S_i = 0$. 
If we assume that a residue is strongly conserved if $S_i \leq 0.1$, then there are very few residues with high sequence conservation. 
These are G15, P21, W22, T35, G43, L54, R57, G95, and G96 (Fig. \ref{fig:conservation}B). 
Sequence entropy is too restrictive in assessing the nature of mutations that are tolerated at a given position.
 The allowed variations in the amino acid substitution is better captured using the chemical sequence 
entropy \cite{Stan_BiophyChem_03}, $S_{CSE} = -\sum_{x=1}^{4}{p_i^xln(p_i^x)}$ where the twenty amino acids are divided into four 
classes, namely, Hydrophobic (H), Polar (P), positively charged (+), and negatively charged (-) \cite{Stan_BiophyChem_03}. Using 
chemical sequence entropy, residues, namely, I14, G15, M20, P21, W22, D27, 
F31, T35, V40, I41, M42, G43, T46, W47, S49, I50, G51, L54, R57, I60, I61, L62, S63, I91, M92, V93, G95, G96, V99, Y100, L110, T113, I115, and F125, are strongly 
($S_{CSE} \leq 0.1$) conserved.

It is not surprising that many of the residues identified in the allostery wiring diagram (Fig. \ref{fig:conservation}) are also strongly conserved 
as they 
are associated directly with the binding surface that stabilize the closed conformation.
The SCA also identifies residues N18, L28, K38, V72, S77, A84, G97, and Q108 that are neither highly conserved nor adjacent 
to conserved residues. It appears that many of the residues in the network are relevant for executing dynamical motions that drive the allosteric transitions in DHFR 
or for cofactor binding.
For example, N18 forms contact with H124 in the CS. 
Similarly, during the OS$\rightarrow$CS transition, 
L28 comes close to I50 upon binding of various ligands which results in the closure of the active site cleft \cite{Kraut_Science_86}. Residue K38, which is in the hinge region, facilitates rotation of the adenosine binding domain towards the loop domain (residues M1-D37 and A107-R159)  \cite{Schnell_Annu_Rev_04}.  

The key residues in the allostery wiring diagram have been shown in previous theoretical and experimental studies to be important either in catalysis or in binding of cofactors. 
Benkovic and coworkers showed that mutations of residues (M42 and G121) that are far from the active site affect the hydride transfer rates \cite{Benkovic_Biochem_97,Benkovic_Philosophical_06}. 
Based on equilibrium covariance matrix fluctuations of the $C_\alpha$ atoms obtained from all atom MD simulations, 
Rod et al showed that interactions of M42 with other residues (H45, D28, S49) would also be involved in the CS$\rightarrow$OS conformational transition \cite{Brooks_PNAS_03}. 
Mutations of positions M42 and/or G121, that lead to anti-correlated motions between the two subdomains,  are found to 
be part of the predicted allostery wiring diagram. Hammes-Schiffer 
used sequence conservation of a small dataset of DHFR sequences to identify a network of residues whose coordinated motion is apparently 
linked to catalysis \cite{Schiffer_PNAS_02,Schiffer_JPCB_02}. 
Among them, I14 is found to be in the allostery wiring diagram that we have identified using SCA. It was also found that motions of residues 
W22, D27, M42, I60, L62, and T113 which forms hydrogen bond network with DHF in the active site might also be involved in coupled promoting motions \cite{Schiffer_PNAS_02,Schiffer_JPCB_02}. 
Taken together, the present and previous studies show that the allostery wiring diagram, that represents the network of signaling residues in DHFR, is 
spread throughout the structure (Fig. \ref{fig:conservation}). More importantly, many non-conserved residues are part of the network. 
     
{\bf Motions between the two subdomains in the CS and OS states are anticorrelated}

The Root Mean Square Deviation (RMSD) between the closed and occluded crystal structures of \textit{E. Coli.} DHFR is only 1.18 $\AA$. However, the RMSD of the active Met20 loop (residues A9-L24) between the two end point structures is alomost three times larger ($\approx$ 3.35 $\AA$).  In order to assess the differences in the structures of the two states at finite temperature, we equilibrated the OS and CS conformations at 300 K.  Comparison of the thermally averaged contact maps  shows that the closed state differs from the occluded conformation mainly in the Met20 loop and the secondary structural elements that are affected by the motions (see below) of the Met20 loop (data not shown).  The largest changes occur in the $\beta$F-$\beta$G (D116-D132) and  $\beta$G-$\beta$H loop (D142-S150) loops, and the $\alpha$-helix H2 of the adenosine binding domain(residues R44-I50). The crystal structures and the thermally equilibrated CS and OS states also show that, in the CS$\rightarrow$OS transition the conformational fluctuations in the Met20 loop have to be accompanied by the following changes: (1) Contacts between the Met20  and $\beta$F-$\beta$G loops should be ruptured. (2) Interactions between helix 2 and the Met20 loop should be disrupted, and reform in a different location; (3) Stabilizing contacts between Met20 and $\beta$G-$\beta$H loops should form.  If these processes are disrupted then it is likely that the catalytic efficiency of DHFR may be compromised. Indeed, experimental findings of the importance of mutating M42, G121, S148 or any two of these residues on the hydride transfer rates can be rationalized based solely on a static picture \cite{Brooks_PNAS_03}. 

The correlated motions in DHFR are computed using time average covariance matrix defined as,

\begin{equation}
\langle C_{ij}(X)  \rangle=\frac{1}{T_{obs}} \int_0^{T_{obs}} {\Delta \hat r_i(t)} \cdot {\Delta \hat r_j(t)} dt
\label{eq:correlation}
\end{equation}
where  $\Delta \hat r_i(t)=(\vec r_i(t)-\vec r_i^{wt})/|\vec r_i(t)-\vec r_i^{wt}|$ is the unit vector of the displacement of the $i^{th}$   $C_{\alpha}$ atom with respect to its initial value, and $X$ is either CS or OS.  The direction of motion of the $i^{th}$ residue is given by $\Delta \hat r_i$. If $\langle C_{ij} \rangle$ is positive then the motion of the two residues $i$ and $j$ are correlated while negative values correspond to anti-correlation.  For perfectly correlated (anti-correlated) residues $\langle C_{ij} \rangle$ is +1 (-1).   The covariance map for the CS shows anti-correlated motion between the Met20 loop and the adenosine binding domain, as well as between the $\beta$F-$\beta$G and $\beta$G-$\beta$H loops (see the dark blue regions in Fig. \ref{fig:correlation} ).  Thus, the adenosine  binding domain and the loop domains move in an anti-correlated manner.  Similar conclusions were obtained using all atom simulations of the WT DHFR \cite{Brooks_PNAS_03}. The present simulations and previous MD studies \cite{Brooks_PNAS_03,Brooks_Proteins_04,Schiffer_PNAS_02,Schiffer_JPCB_02,Schiffer_JPCB_04,Schiffer_PNAS_05} point to the importance of correlated motions between regions that are spatially well separated. The cross correlations in the inter-domain motions shown in Fig. \ref{fig:correlation} are obtained by averaging the structural fluctuations over 0.1 ms, and may well be relevant in facilitating cofactor binding and solvent rearrangement needed for catalysis \cite{Warshel_Biochem_07}. The static picture alone is not sufficient to describe the kinetics of transitions between the CS and OS.  Only by probing  the kinetics of conformational fluctuations in the CS$\rightarrow$OS (and the backward) transitions  we can predict the order of events that results in the conformational changes in the all important Met20 loop.

The bottom panel in Fig. \ref {fig:correlation} shows the differences in the covariance matrices  $\Delta C_{ij}= \langle C_{ij}(CS )\rangle - \langle C_{ij}(OS) \rangle$ in regions that are significantly different between the two allosteric states.  The red region between the Met20 loop and the adenosine binding domain indicates more anti-correlated motions in OS than in CS. The blue region between the Met20 loop and the other two loops shows less anti-correlated motions and more correlated motions in OS than in CS.  
 
{\bf Kinetics of CS$\rightarrow$OS transition involves deformation of the Met20 loop}

In order to dissect the kinetics of structural changes in the Met20 loop during the forward (CS$\rightarrow$OS) and the backward (CS$\rightarrow$OS) directions we have performed a number of simulations using the procedures described in the Methods section.  
Although it is clear that Met20 loop plays essential role in 
this transition, the order of events that drives its conformational change is not known \cite{Schnell_Annu_Rev_04,Boehr_Science_06}. We monitor  the Met20 loop kinetics using two surrogate reaction coordinates. One is the global RMSD, $\Delta_G$, that is obtained by aligning the instantaneous conformation of DHFR at time $t$ either with the CS or the OS structure. During the CS$\rightarrow$OS transition, $\Delta_G$,  should increase with respect to the CS and decrease with respect to OS.  From the time dependent variations in $\Delta_G$ we can infer the changes in the Met20 loop with respect to the entire structure.  To determine the changes that are localized in the Met20 loop we calculated a local RMSD, $\Delta_L$, which uses only the coordinates of the active loop. From the time dependent changes in $\Delta_L$, which is computed by aligning the Met20 loop and computing its RMSD (with respect to the starting conformation) during the two transitions, we can explicitly identify the dominant motions (translation, rotation, or twist) of the loop. The kinetics expressed in terms of the local coordinate $\Delta_L$ yields the conformational changes of only the Met20 loop. 

The time dependent changes in the  global RMSD, $\Delta_G(t)$, with respect to the CS show considerable dynamical heterogeneity (Fig. \ref{fig:multiple_crossing}).  The bottom panel in Fig. \ref{fig:multiple_crossing}, for one trajectory, shows that there are multiple recrossings across the transition region which is suggestive of a rather broad transition region (see below).  Prior to the CS $\rightarrow$ OS transition ($t < 80 \mu s$ in Fig. \ref{fig:multiple_crossing}) $\Delta_G(t)$ undergoes substantial fluctuations which suggests that high energy states are being sampled while DHFR is in the CS. More importantly, such fluctuations can lead to infrequent visits to conformations that are similar to the OS state (see below).  The broad distribution of transition times and multiple recrossings attests to the plasticity of the enzyme during the conformational transition.

Although there is great diversity in the dynamics of the individual trajectories $<\Delta_G(t)>$ and $<\Delta_L(t)>$, obtained by averaging over an ensemble of initial conformations, can be approximately described using a two-state model (Fig. \ref{fig:forward_transition} A). Comparison of $\Delta_G(t)$ and  $\Delta_L(t)$ shows (Fig. \ref{fig:forward_transition}A) that deformations of the Met20 loop occurs after the  global motions in the CS$\rightarrow$OS transition.  Because the long time values of $<\Delta_L>$ are less than $<\Delta_G>$ values, we surmise that the structural changes in the Met20 loop involve translation and rotational motion towards the OS structure.

{\bf Sliding of the Met20 loop across $\alpha$2 is the rate limiting step in the CS$\rightarrow$OS transition} 

In order to understand the mechanism of the communication during the CS$\rightarrow$OS transition we monitored the local movements of the Met20 loop and the helix $\alpha$2 in the adenosine binding domain. Rupture of the  contacts in the CS state (Asn18-His45,  Asn18-Ser49, and Ala19-Ser49) and formation of Glu17-Ser49 during the CS$\rightarrow$OS transition facilitates the sliding of Met20 along $\alpha$2 (Fig. \ref{fig:contacts}A). The relative sliding motion between $\alpha$2 and the Met20 loop enables NADPH to move closer to DHF.

In the loop subdomain, the flexible Met20 loop interacts simultaneously with both the $\beta$F-$\beta$G and $\beta$G-$\beta$H loops \ref{fig:catalysis_cycle}. In order dissect the order of events that occurs in the CS$\rightarrow$OS transition we have computed the kinetics of breakage and formation of a number of contacts involving the two loops (Fig. \ref{fig:contacts}A-C ). By fitting the time-dependent changes in the formation and rupture of contacts to single exponential kinetics we find that the rupture of contacts between Met20 loop and $\beta$F-$\beta$G loop in CS as well as formation of contacts between residues in the Met20 loop and $\beta$G-$\beta$H loop occur nearly simultaneously (on the $\mu s$ time scale) (see (Fig. \ref{fig:contacts}A-C)). Only subsequently (on a time scale of about 2 $\mu s$), the interaction between Glu17 (in Met20 loop) and Ser49 (in $\alpha$2) that exists only the CS state, takes place. Thus, the sliding of Met20 loop on $\alpha$2 requires coordinated motion of a number of residues in the loop domain.

We can further dissect the nature of the sliding motion of the Met20 loop along $\alpha$2 by simultaneously measuring the changes in the angles and the distances between selected residues. We have computed the time-dependent changes in the distances between Asn18-His45 ($R_1$), Ala19-Ser49 ($R_2$), Asn18-Met42 ($R_3$) and Glu17-Ser49 ($R_4$), respectively. The sliding motion is vividly illustrated using the changes in the angles that the vectors $\vec R_1$, $\vec R_2$, $\vec R_3$ and $\vec R_4$ make with the axis of $\alpha$2. Angles are defined as $\alpha_i=cos^{-1}(\hat R_i \cdot \hat U_{H2})$ $i=1,2,3,4$ and $\hat U_{H2}$ is the unit vector of the $\alpha$2 helix axis. The two-dimensional projection of ($R_i, \alpha_i$) ($i=1,2,3,4$), that represents the values of ($R_i, \alpha_i$) that are sampled in the kinetic trajectories, shows that $\alpha_i$ values decrease monotonically during the CS$\rightarrow$OS transition. The averages over all the trajectories for $\alpha_i$ also show a monotonic decrease. The averages also show that the $\alpha_i$ values are either clustered around the CS or the OS state. In other words, there is very little backtracking in the sliding movement of the Met20 loop along H2. The histogram of the angels and distances sampled during the transition in Fig. \ref{fig:R_alpha} also shows the fluctuations in $\alpha_i$ ($i=1,2,3,4$) are centered around the OS values which suggests that (in terms of these microscopic variables) that the transition occurs when the conformation is close to the OS state. This result is also in accord with the results in Fig. \ref{fig:contacts}A which shows that Glu17-Ser49 only forms when the interactions in the CS are ruptured, a process that occurs closer to the completion of the CS $\rightarrow$ OS transition. The structural changes that accompany the sliding motion of the Met20 loop involves concerted motion of a number of residues (see the diagram on the left in Fig. \ref{fig:Motion}). The figures summarize the collective motions of residues in both the subdomains that facilitate the structural deformations in the Met20 loop.

{\bf Cysteine crosslink inhibits CS$\rightarrow$OS transition}

The kinetics in both the forward and the backward (see below) transitions show that the coordinated motion in the loop subdomain plays an important role in enabling the Met20 loop communicate with adenosine binding domain. In the crystal structure of CS, the distance between the $C_{\alpha}$ atoms of Met16 and Gly121 is about 4.3$\AA$. It is possible to mutate these residues to Cys to establish a disulfide cross link. We have simulated the kinetics of the CS$\rightarrow$OS transition in the crosslink mutant (referred to as CL) to assess the extent to which the motion of the Met20 loop is inhibited. Previously, it has been argued that constraining even residues that are 28 $\AA$ apart can affect hydride transfer rates \cite{Schiffer_JPCB_06_Freezing_Motion}. Our purpose in studying the CL mutant is to see how the strain in the loop domain would affect the communication between the two domains. Since the disulfide bond constrains the distance between Met16 and Gly121 to 4.3 $\AA$, the anti-correlated motion between Met20 loop and $\beta$F-$\beta$G loop should be impeded. The time dependencies of $<\Delta_L(CS|t)>$ and $<\Delta_L(OS|t)>$ show that the Met20 loop does not fully adopt its conformation in the OS state (compare Fig. \ref{fig:forward_transition}A and B). Similarly, the long time values of $<\Delta_G(CS)>$ and $<\Delta_G(OS)>$ in the mutant are different than in the WT (see Fig. \ref{fig:forward_transition}B). In the WT, $\beta$G-$\beta$H loop are involved in the coordinated motion between two domains. Surprisingly, the crosslink has little effect on the relative motion between $\beta$G-$\beta$H loop and the Met20 loop. The time dependent changes that monitor the formation of contacts between Trp22-His149 and Asn23-His149 are similar in the WT and the crosslink mutant. Because the interactions between the Met20 loop and $\beta$G-$\beta$H loop are not fully inhibited in the CL, the sliding motion across $\alpha$2 with the formation of Glu17-Ser49 can occur (Fig. \ref{fig:contacts_CL}A) albeit less efficiently. We predict that due to the incomplete CS$\rightarrow$OS transition the crosslink will dramatically affect the rate of the forward hydride transition. Experiments using CL can shed further light on the importance of enzyme motion in catalysis which still remains controversial \cite{Schiffer_PNAS_02,Schiffer_JPCB_02,Warshel_ChemRev_06}.

{\bf Deformation of the Met20 loop drives the global motion during the OS$\rightarrow$CS transition}

The time constant for the local kinetics of the Met20 loop in the OS$\rightarrow$CS transition obtained from $<\Delta_L(t)>$ (Fig. \ref{fig:reverse_transition}) is greater than the time scale in which $<\Delta_G(t)>$ changes. This implies that the Met20 gliding across $\alpha$2 is the first event in the OS$\rightarrow$CS transition. In contrast, during the CS$\rightarrow$OS transition, only in the final stages does the Met20 loop occludes the active site.

Although the initial change in the OS$\rightarrow$CS transition involves the deformation of the Met20 loop (Fig. \ref{fig:reverse_transition}) the microscopic events that drive this transition are distinct from those seen in the CS$\rightarrow$OS transition. Remarkably, the rupture of Glu17-Ser49 occurs only after the formation of the contact between the Met20 loop and $\alpha$2. The time dependent changes in the contacts present only in the CS state (Asn18-His45, Asn18-Ser49, and Ala19-Ser49) occur while Glu17-Ser49 (in the OS state) contact still persists (Fig. \ref{fig:contacts_CL}B). We suggest that binding of NADPH, which is required for THF to be released, assists in the formation of contacts between Met20 loop and $\beta$F-$\beta$G, and between Met20 loop and $\alpha$2. Only after these contacts are established the contact between Glu17-Ser49 ruptures (Fig. \ref{fig:contacts_CL}B). Upon rupture of the Glu17-Ser49 contact, the Met20 loop slides back to its closed conformation, and THF is released.

The simulations also show coordinated motions among the three loops in the loop subdomain during the OS$\rightarrow$CS transition (see the right side of Fig. \ref{fig:Motion}). From the analysis of the time-dependent changes in the distances between a number of residues we conclude that $\beta$F-$\beta$G loop stretches the Met20 loop by forming a number of contacts (Gly15-His124 ,Met16-Glu120, Met16-Gly121, Met16-Asp122, Met16-Thr123, Glu17-Gly121, and Glu17-Asp122) with the Met20 loop. In concert with these events the strain imposed on the Met20 loop by the $\beta$G-$\beta$H loop is released by rupture of contacts (Trp22-Ser148, Trp22-His149, and Asn23-His149) with the Met20 loop. The pull (by the $\beta$F-$\beta$G loop) and push (by the $\beta$G-$\beta$H loop) action on the Met20 loop must take place before the Met20 loop slides back to its conformation on the CS state (Fig. \ref{fig:contacts}B). These results show that the pathways in the OS$\rightarrow$CS transition are not the reverse of what transpires during the CS$\rightarrow$OS transition. The structural changes in the Met20 loop and the concerted motions of a number of residues that drive these changes are shown on the right side of Fig. \ref{fig:Motion}.

{\bf Residues in the allostery wiring diagram code for ligand binding and dynamics}  

The SCA predicts a number of residues that are expected to be relevant either in the motion of DHFR or in the function (Fig. \ref{fig:conservation}A). Some of the residues in the network are related to cofactor binding and interaction with the active site while others are directly involved in accommodating the motion of the Met20 loop during the CS$\rightarrow$OS transition. For example, SCA identified Leu28, Ala29, and Ser63 (Fig. \ref{fig:conservation}) all of which are involved in ligand binding or binding-involved dynamics. The amino acid at location 29, which in \textit {E. Coli} DHFR is Ala, is in contact with His28 show isomerization between two isoforms of the apoenzyme \cite{Schnell_Annu_Rev_04}. In \textit {L. casei} enzyme the conversion between the isoforms occurs only for the folate-bound complex while in human DHFR there appears to be only conformation in the methotrexate (MTX) DHFR complex \cite{Schnell_Annu_Rev_04}. The importance of Ser63 in maintaining hydrogen bond with NADPH was noted in the molecular dynamics simulations \cite{Schiffer_PNAS_02,Schiffer_JPCB_02}. Similarly, Asp27 is involved in hydrogen network with DHF in the active site \cite{Kraut_Science_86}. The network predicted by SCA also contains Ile60 and Leu62 both of which have been recognized to be dynamically involved in interactions with Met20 loop. SCA also suggests that Ile94 and Gly97 should play a role in the function of DHFR. Because SCA cannot assess the importance of absolutely conserved residues it is likely that neighboring residues Gly95 and Gly96 may be relevant in the reaction cycle of DHFR \cite{Schiffer_PNAS_02,Schiffer_JPCB_02}. It is noteworthy that the SCA identified a network of residues in the helical region $\alpha$2 in the adenosine binding domain as being important. The present simulations show that the critical sliding motion of the Met20 loop along $\alpha$2 completes the allosteric transitions. Mutations in the region (Ile41-His45), that is far from the active site, have great influence on the forward hydride transfer reaction without affecting cofactor binding \cite{Schiffer_JPCB_04,Schiffer_PNAS_05}. It appears that the predictions of the SCA can be rationalized in light of a number of experimental and theoretical studies that have identified the importance of concerted motions among a sparse network of residues on the reaction cycle of DHFR. The sequence-based approach fails to identify key residues (Gly121 being the most important) which apparently plays a role in catalysis \cite{Benkovic_Biochem_97}. 

{\bf The average transition state structure resembles OS (CS) in the CS$\rightarrow$OS (OS$\rightarrow$CS) transition. }

We have used the global RMSD ($\Delta_G$) as a surrogate reaction coordinate to determine the structures of the transition state ensemble (TSE). We assume that the transition state (TS) for a molecule  is reached for the first time at $t_{TS}$, if $|\Delta_G^CS(t_{TS})-\Delta_G^OS(t_{TS})|<\epsilon(=0.5\AA)$ is satisfied.  Our  criterion places the transition state equidistant (in terms of the global RMSD) from the CS and OS. Comparison of the contact maps (data not shown) for the TSE, CS, and OS shows that both the transitions exhibit major changes more with respect to the starting than the ending state.  The largest changes between the CS and OS states, which take place in the Met20 and $\beta F-\beta G$ loops, occur before the transition state is reached.

The heterogeneity observed in the dynamics of the CS$\rightarrow$OS transition is also reflected in the distribution $P(t_{TS})$ of the transition time $t_{TS}$ (Fig. \ref{fig:TS_distribution}A). Surprisingly,  $P(t_{TS})$ is approximately uniform in the CS $\rightarrow$ OS transition (Fig. \ref{fig:TS_distribution}A).  As a result, the TSE structures are much less heterogeneous in the forward than in the backward direction (Fig. \ref{fig:TS_distribution}C). However, the spread in $t_{TS}$ is broader in the forward direction compared to the backward direction. From the TSE we can compute a Tanford $\beta$-like parameter, $q^\ddagger$, ($0 \le q^\ddagger \le 1$) using 
\begin{equation}
q^\ddagger=\frac{max(\Delta^\ddagger)-\Delta^\ddagger}{max(\Delta^\ddagger)-min(\Delta^\ddagger)}
\label{eq:qdagger}
\end{equation} 
where $\Delta^\ddagger=(\Delta_G^C(t_{TS})+\Delta_G^O(t_{TS}))/2$, $max(\Delta^{\ddagger})$ and $min(\Delta^{\ddagger})$ are the maximum and minimum values of $\Delta^{\ddagger}$ respectively. If $q^{\ddagger}$ is close to 0 (1) then the most probable TS is starting (ending) allosteric state. For the CS $\rightarrow$OS transition the average value of $q^{\ddagger}$ is 0.66 (Fig. \ref{fig:TS_distribution}B) which implies that the TSE structures are more OS-like (see the average TS structures in Figs. \ref{fig:forward_transition} and \ref{fig:reverse_transition}). Although the distribution $P(q^{\ddagger})$ for the OS$\rightarrow$CS is very broad (Fig. \ref{fig:TS_distribution}D) the most probable $q^{\ddagger}$ is closer to the CS than to the OS.  Thus, in both the transitions the average TSE structures resemble the high energy allosteric states.  This observation supports the recent inferences drawn from the NMR relaxation time measurements \cite{Boehr_Science_06} that the high energy conformation is populated (in the pre-equilibrium sense) in the both the allosteric states. From the Hammond postulate it follows that the TSE structures should resemble the high free energy states in accord with the present simulation results.  Surprisingly, we further predict that the TSE structures for the OS$\rightarrow$CS transition is conformationally much more heterogeneous than in the forward direction (Compare Figs. \ref{fig:TS_distribution}B and \ref{fig:TS_distribution}D). 

{\bf A small fraction of OS (CS) is present under equilibrium conditions in the CS (OS) state}

Although the dominant basin of attraction corresponds to a unique native folded state enzymes can sample other conformations, albeit not frequently, through thermal fluctuations. Some of the conformations that are sampled in the ensemble of the equilibrated CS can correspond to the structures in the OS \cite{Boehr_Science_06}. Allosteric mechanism based on the preexisting equilibrium \cite{Tsai_ProtSci_99} is qualitatively different from the induced-fit model \cite{Koshland_PNAS_58} which posits that the conformational transitions in the CS state occur only after the ligand binds. Indeed, several experiments, including the recent reports on DHFR \cite{Boehr_Science_06}, suggest that a small population of OS conformations are in equilibrium with the CS structures. Similarly, we expect that CS structures should be accessible when the molecule is predominantly in the OS.

In order to probe the validity of the conformational selection model \cite{Tsai_ProtSci_99} we calculated the distribution $P(\Delta \Delta_G)$ where 
\begin{equation}
\Delta \Delta_G = \Delta_G^{OS} - \Delta_G^{CS}
\label{eq:ddg}
\end{equation}
where $\Delta_G^{OS}$  is the equilibrium RMSD of conformations in the CS with respect to the OS structure, and $\Delta_G^{CS}$ is the corresponding RMSD with respect to the CS structure. If DHFR is in the CS without ever sampling the OS-like structures then we expect that $\Delta_G^{CS} \approx 0$.  As a result,  $P(\Delta \Delta_G)$ should be identically zero whenever $\Delta \Delta_G < 0$. Thus, the observation of negative values of $\Delta \Delta_G$ is an indication of preexisting OS-like structures even under equilibrium conditions that favor CS (Fig. \ref{fig:pre_eq}A). Figure \ref{fig:pre_eq}A (Fig. \ref{fig:pre_eq}B) shows that $P(\Delta \Delta_G)$ is non-zero for a small range of negative (positive) $\Delta \Delta_G$ in the ensemble of CS (OS). The population of the micro species CS (OS) in the OS (CS) basin is $\sim 3\%$ ($\sim 1\%$). Surprisingly, these estimates are similar to the values reported by Boehr \textit{et al.} \cite{Boehr_Science_06}. The presence of higher energy species also suggests, in accord with the Hammond postulate that the TS structure should be OS-like in the CS$\rightarrow$OS transition. This inference, which follows from the conformational selection model is in accord with our simulations. We predict that mutations that destabilize either the CS or the OS will affect the kinetics of the allosteric transitions.

{\bf Concluding Remarks}

From the perspective of allostery, it is not surprising that communication between residues that are spatially well separated facilitates the CS$\rightarrow$OS transition \cite{Changeux_Science_05}. We have used sequence-based method to identify a network of mechanically important residues that could control the kinetics of conformational transitions. The residues in the network are dispersed both in the adenosine binding and the loop domains. Coordinated motions among these residues and others control the structural transitions, and perhaps the forward and backward hydride transfer reaction.  Surprisingly, several of these residues are not strongly conserved although their chemical character are often preserved across the various species. Hydride transfer experiments on the wild type DHFR and its mutants \cite{Schiffer_Benkovic_Annu_Rev_06,Benkovic_Biochem_02} have already pointed out the importance of many of the residues in the network. In addition, all atom molecular dynamics simulations \cite{Schiffer_PNAS_02,Brooks_PNAS_03} and NMR experiments \cite{Schnell_Annu_Rev_04,Boehr_Science_06} have implicated the key role of the network residues in the dynamics of DHFR. Although it is difficult to unambiguously establish a direct link between the DHFR motions (equilibrium or dynamic) and hydride transfer reaction \cite{Warshel_ChemRev_06}, the perturbation of these residues will affect magnetic resonance relaxation dispersion.

The kinetics of the allosteric transitions in the forward (CS$\rightarrow$OS) and the reverse (OS$\rightarrow$CS), using the SOP model, reveal in great detail the order of events that results in the movement  of the Met20 loop. In the forward direction, several contacts in the CS state rupture and new ones form in the OS. The concerted kinetics associated with these contacts, most of which are associated with the Met20, $\beta$F-$\beta$G, and $\beta$G-$\beta$H loops facilitate the sliding motion of the Met20 loop so that it occludes the active site (Fig. \ref{fig:catalysis_cycle}). Surprisingly, the pathways in the OS$\rightarrow$CS transition are not the reverse of the forward reaction. In particular, the interactions between Glu17-Ser49, whose rupture facilitates the sliding of the Met20 back to its CS position, persist till late in the OS$\rightarrow$CS transition. In the forward direction, Glu17-Ser49 contact occurs late for the sliding motion of Met20 along $\alpha$2 to take place. The broad transition state region, both in the forward and backward directions,  attests to the inherent plasticity of enzymes in general, and DHFR in particular.  These results support the notion that mutations that inhibit the equilibrium fluctuations leading to the population of the minor species can adversely affect the rates of hydride transfer reaction.  Indeed, the observed decrease in the hydride transfer rate in G121V has been rationalized using this picture \cite{Benkovic_Biochem_97}.

Of particular importance is the link between the present studies and the recent NMR relaxation measurements Boehr \textit{et al.} \cite{Boehr_Science_06} which showed that, at equilibrium, there is a small percentage of OS structures in the ensemble of CS conformations. Similarly, when DHFR is in the OS state dynamical fluctuations populate a small ($\sim (1-3)\%$) of CS structures. Our simulation results are in accord with the NMR experiments  \cite{Boehr_Science_06}. These results support the emerging notion that in enzymes conformations resembling the cofactor-bound structure is already present in the apoenzyme. The cofactors dynamically funnel the minor populations so that the equilibrium shifts to the haloenzyme. The present simulations show that such conformational fluctuations occur on $\mu$s time scale. Because of the simplicity of the SOP model the estimated time scale should be taken as a lower bound. The ability to access the higher free energy states on ($\mu{s}$-$ms$) time scale is a consequence of the conformational heterogeneity of the enzyme which leads to low barriers separating the relevant kinetic states. In DHFR, this is reflected in the broad TSE with heterogeneous structures that results in a broad distribution of crossing times between the allosteric states.

We also obtained the temperature ($T$) dependence of the rates of the forward and reverse transition for the WT and 
the forward transition for the CL. The rates were computed by fitting the 
time dependence of $<\Delta_G(t)>$ for $T$ in the range $285K<T<315K$. The averaging is performed using 
20 trajectories. We find that the three rates follow the Arrhenius behavior. Because the SOP is a coarse-grained 
model the activation barrier is severely underestimated. Nevertheless, the results show that the rates of the allosteric 
transitions are enhanced as $T$ increases. Needless to say that altering $T$ might also change the reorganization free 
energies of the solvent which could be a dominant factor in determining the catalytic rates \cite{Warshel_ChemRev_06, Warshel_Biochem_07,Warshel_JPCB_07}.

The predicted temperature dependence of the CS$\rightarrow$OS transition might provide a way to test the extent to which correlated enzyme dynamics is important in catalysis. If the rates of conformational fluctuations are drastically different from the rate of the chemical step then it is unlikely that the correlated motions of the enzyme is crucial to catalysis. On the other hand, if the two states are comparable then the conformational changes during the CS$\rightarrow$OS transition might be coupled to hydride transfer. In general, it is important to formulate a method that couples the dynamics of the CS$\rightarrow$OS transition (including the cofactors) and the chemical step. Such a formalism must account for both the kinetics of transitions along the lines described here and the hydride transfer processes described by others \cite{Warshel_Biochem_07,Schiffer_PNAS_02}.

The SOP model, which was introduced to carry out simulations of large systems \cite{Hyeon_Struct_06,Hyeon_PNAS_06}, does not include a number of relevant interactions. Most notably, the lack of explicit model for hydrogen bonds,  prevents us from examining their role in the allosteric transitions. The role the network of hydrogen bonds of DHFR plays in affecting the CS$\rightarrow$OS transition can only be vicariously gleaned using the SOP model. On the other hand, the major advantage of the SOP model is that long time simulations for a large number of trajectories can be carried out. Indeed, the non-trivial prediction that the coordinated motions of specific residues throughout the structure trigger the movement of the Met20 loop is amenable to experimental tests.  The non-conserved residues identified in this work can form the basis of future mutagenesis experiments. We believe that a combination of computational methods (sequence-based technique, coarse-grained and all atom MD simulations), and NMR, single molecule, and biochemical experiments are needed to fully dissect the interplay between enzyme motion and catalysis. 

{\bf Methods}

{\bf Statistical coupling analysis:}

In order to identify the residues that are involved in transmitting allosteric signals, we use our formulation \cite{Dima_ProtSci_06} of the Sequence-based Statistical Coupling 
Analysis (SCA) introduced by Lockless and Ranganathan \cite{RR_NSB_03,RR_PNAS_03,RR_PNAS_04}. A statistical free energy-like function at each position, $i$, in a multiple sequence alignment (MSA) is defined as

\begin {equation}
\frac{\Delta G_i} {k_{B}T^*}=\sqrt{\frac{1}{C_i}\sum_{x=1}^{20}{[p_i^xln(\frac{p_i^x}{p_x})]^2}}  
\label{eq:delta_Gi}
\end {equation}
where, $k_{B}T^*$ is an arbitrary energy unit, $C_i$ is the number of types of amino acid that appears at position $i$, $p_x$ is the mean frequency of 
amino acid $x$ in the MSA. In eq.(1) $p_i^x=\frac{n_i^x}{N_i}$, where $n_i^x$ is the number of times amino acid $x$ appears at position $i$ in the MSA, 
and $N_i=\sum_{x=1}^{20}n_i^x$.

The basic hypothesis of the SCA is that correlation or covariation between two positions $i$ and $j$ may be inferred by comparing the statistical properties of the MSA and a 
subalignment of sequences (derived from the MSA) in which a given amino acid is conserved ($S_j=0$) at $j$. 
The restriction that $S_j=0$ in the subalignment is referred to as sequence 
perturbation at position $j$. The effect of perturbation is assessed using,

\begin {equation}
\frac{\Delta \Delta G_{ij}}{kT^*}=\sqrt{\frac{1}{C_i}\sum_{x=1}^{20}{[ p_{i,R}^{x}ln(\frac{p_{i,R}^{x}}{p_x})-p_i^xln(\frac{p_i^x}{p_x}]^2}}
\label{eq:delta_delta_Gij}
\end{equation}
where $p_{i,R}^x=n_{i,R}^x/N_{i,R}$, $n_{i,R}^x$ and $N_{i,R}$ are the number of sequences in the subalignment in which $x$ appears in the $i^{th}$ 
position and $N_{i,R}=\sum_{x=1}^{20}{n_{i,R}^{x}}$.

In order to obtain statistically meaningful results using the SCA, it is crucial to choose 
the subalignments appropriately. Let $f=p/N_{MSA}$ where $p$ is the number of sequence in the subalignment and $N_{MSA}$ is the total number of 
sequences in the MSA. We choose $f$ ($=0.35$ for the DHFR family) to satisfy the central limit theorem \cite{Dima_ProtSci_06} so that the statistical properties from the 
subalignments coincide with the full MSA. Using $f=0.35$, we calculated the matrix elements 
$\Delta\Delta G_{ij}$ which estimate the response of position $i$ in the MSA to all allowed perturbations at $j$ ($S_j=0$). The rows (labeled $i$) 
in $\Delta\Delta G_{ij}$ correspond 
to positions in the MSA. We determined the network of covarying residues using the elements 
$\Delta\Delta G_{ij}$ in conjunction with coupled two-way clustering algorithm \cite{Domany_PNAS_00}. The extent to which the rows 
$\Delta\Delta G_{ij}$ and $\Delta\Delta G_{kj}$ are similar is assessed using the Euclidean measure \cite{Dima_ProtSci_06}.  
Because $\Delta\Delta G_{ij}=0$ for perfectly conserved positions and for sites where the amino acids are found at their mean frequencies in the MSA 
($p_i^x=p_i$), the SCA cannot predict the role these residues play in the function or dynamics of the enzyme.  

{\bf Self-organized polymer (SOP) model for  closed (CS) and Occluded (OS) states}

We have carried out Brownian dynamics simulations to obtain the kinetics of transitions between the allosteric states. Because of long-time scales involved in these transitions, we use a 
coarse-grained self-organized polymer (SOP) model for DHFR in the CS and OS states. The validity of the SOP model has  been established in the context of mechanical unfolding of RNA and proteins \cite{Hyeon_BiophyJ_07}, allostery in GroEL \cite{Hyeon_PNAS_06}, and conformational changes in kinesin \cite{Hyeon_PNAS_07}. In the SOP model, the structure of a protein is represented using only the $C_{\alpha}$ coordinates. 
Because we are interested in the kinetics of CS$\rightarrow$OS and OS$\rightarrow$CS transitions 
the energy functions are state dependent. The state-dependent energy function, in the SOP representation of protein 
structures in terms of the $C_{\alpha}$ coordinates $r_i (i=1,2,...N$ with $N$ being the number of amino acids), is 

\begin {eqnarray}
\label{eqn:potential}
V\{r_i|X\}&=&V_{FENE}+V_{NB}^{A}+V_{NB}^{R}\\\nonumber
       &=&-\sum_{i=1}^{N-1}{
                           \frac{k}{2} R_0^2 
                           \log(1-\frac{(r_{i,i+1}-r_{i,i+1}^0(X))^2} 
                                        {R_0^2})
                           } \\ \nonumber
       &+& \sum_{i=1}^{N-3}\sum_{j=i+3}^{N}\epsilon_h[(\frac{r_{ij}^0(X)}{r_{ij}})^{12}-2(\frac{r_{ij(X)}^0}{r_{ij}})^6]\Delta_{ij} \\ \nonumber
       &+& \sum_{i=1}^{N-2}\epsilon_l(\frac{r_{i,i+2}^0}{r_{i,i+2}})^{6} +\sum_{i<j}\epsilon_l(\frac{\sigma}{r_{ij}})^{6}(1-\Delta_{ij}),
\end{eqnarray}
where the label $X$ refers to the allosteric state CS or OS. In Eq. \ref{eqn:potential},  $r_{i,i+1}$ is the distance between two consequentive $C_\alpha$-atoms, and 
$r_{i,j}$ is the distance between the $i^{th}$ and $j^{th}$ $C_{\alpha}$ atoms, and the superscript $0$ denotes their values in the crystal structure of 
the allosteric state $X$.
 The first term in Eq. \ref{eqn:potential}, the finite extensible non-linear elastic (FENE) potential, accounts for chain connectivity. The stability of the state $X$ is described by the 
non-bonded interactions (second term in Eq.\ref{eqn:potential}) that assign attractions between residues that are in contact in $X$. Non-bonded interactions 
between residues that are not in contact in $X$ are taken to be purely repulsive (third term in Eq.\ref{eqn:potential}). The value of $\Delta_{i,j}$ is 1 if 
$i$ and $j$ are in native contact, and is zero otherwise. 

In the SOP model, there are only two independent parameters. The results are insensitive to the precise values of $k$, $R_0$, and $\epsilon_l$. The two key parameters are 
$\epsilon_h$, a single energy scale that describes the stability of state $X$, and the cutoff
distance $R_C$  for native contacts. Because $R_C$ is, to a large extent, determined by the contact map, there is very little freedom in its choice. We assume that native contact exists 
if the distance between the $i^{th}$ and $j^{th}$ $C_\alpha$ atoms is less than 8 $\AA$.  
The spring constant $k$ in the FENE potential (first term in Eq.\ref{eqn:potential}) for stretching the covalent bond is 
20 kcal/(mol$\AA^2$), and the value of $R_0$, which gives the allowed extension of the covalent bond, is 2 $\AA$. 

{\bf Brownian dynamics simulations of allosteric transitions}

The kinetics of forward and backward transitions between CS and OS are probed using a method that was recently used to study allosteric transitions  in GroEL \cite{Hyeon_PNAS_06}. 
The basic assumption of the method is that the local strain that DHFR experiences (due to a ligand or cofactor binding) propagates faster across the structure than the 
conformational transitions that lead to the CS$\rightarrow$OS and vice versa. In other words, the global structural relaxation is the slow step in the allosteric transition. 

Using the SOP model,
we simulated the transition between the CS (PDB code 1RX2) and
OS states (PDB code 1RX7) by assuming that the dynamics of the protein
can be adequately described by the Langevin equation in the overdamped limit.  Transitions 
from one allosteric state (CS) to another (OS) are induced by starting from a conformation corresponding to the CS. The transition is induced by using 
the forces computed from 
the OS, i.e. from $H\{r_i|OS\}$. The explicit equations of motion for CS$\rightarrow$OS are,
\begin{equation}
\zeta \frac{\partial r_{i}}{\partial t} = - \frac{\partial H(r_{i}|OS)}{\partial 
r_{i}} + F(t)
\label{eqn:Langevin_eq_motion}
\end{equation}
where $H(r_{i}|OS)$ is the  Hamiltonian of the OS state, $\zeta$ is the friction coefficient, $F(t)$ is the random force, and $r_{i}$ is the position
of the $i^{th}$ residue at time  $t$. The initial ($t$ = 0) value of $r_{i}$  is taken from the Boltzmann
distribution at temperature $T$ corresponding to the CS state,
\begin{equation}
P(r_{i}(0)) \sim e^{-\beta H(r_{i}|CS)}
\label{eqn:distrib_ri_zero}
\end{equation}
where $\beta = \frac{1}{k_{B}T}$, and  $k_{B}$ is the Boltzmann
constant. The random force, $F(t)$ satisfies 
\begin{equation}
<F(t)> = 0
\label{eqn:rand_force_av}
\end{equation}
and
\begin{equation}
<F(t)F(t^{'})> = 2 k_{B}T \zeta \delta(t-t^{'}),
\label{eqn:rand_force_var}
\end{equation}
where the averages are over the trajectories. As long as potential conditions are satisfied \cite{Gardinerbook} our method for
inducing transitions ensures that, at long times, DHFR will explore the conformations corresponding to the OS state.  At long times, the
ensemble of conformations obeys the Boltzmann distribution corresponding to the OS state so that  $P(r_{i}|(OS)) \sim e^{-\beta H(r_{i}|OS)}$.  Thus, on general theoretical grounds
our procedure guarantees that CS$\rightarrow$OS transition can be realized, and that the dynamics represents the microscopic events that drive the transition of interest. To monitor the reverse reaction, we begin with an initial equilibrated ensemble of structures corresponding to the OS state and integrate the equations of motion (Eq. \ref{eqn:Langevin_eq_motion}) with the forces arising from the CS state.

The procedure used to induce the allosteric transition hinges on the physically reasonable assumption that the time scales involved in the conformational changes in the enzymes are much longer than the time needed for locally-induced strain (due to cofactor binding for example) to propagate through the structure. In order to test this assumption, we varied the rate at which the CS$\rightarrow$OS transition is allowed to take place by allowing $r^\circ_{ij}$(CS$\rightarrow$OS) to evolve slowly on a relatively long time scale.  We accomplish the slower evolution using,

\begin {equation}
r^\circ_{ij}
(CS\rightarrow OS)=
\frac{
(K-k)r^\circ_{ij}(CS)+kr^\circ_{ij}(OS)}
{K}.
\end{equation}
The majority of the results were obtained using $K=k=1$, To vary the strain propagation time we also performed simulations using $K=100$ and increased $k$ in steps. The CS$\rightarrow$OS conformational switch was made smoothly over a range of times, namely, 0.12$\mu$s, 0.16$\mu$s, 0.2$\mu$s, and 0.24$\mu$s. The kinetics of $<\Delta_G(t)>$ for the CS$\rightarrow$OS transition, that reports on the time-dependence of the global RMSD changes, for the four cases coincide with the results in Fig. \ref{fig:forward_transition} and \ref{fig:reverse_transition} (data not shown). These additional simulations justify the assumption underlying our procedure for inducing the conformational transition.

 {\bf Time scales and simulation details:} 

In order to decipher the events that drive the conformational changes in the Met20 loop during the catalytic cycle, we performed three different simulations. We first simulated the kinetics of CS$\rightarrow$OS  transitions to determine the order of events during the forward reaction. In order to assess the roles of the residues G121 and M16 (the $C_\alpha$ distance between the two is about 4.3$\AA$ apart in the CS state), in  the CS$\rightarrow$OS transition we mutated M16-G121 interaction by a disulfide bond. Such a cross link, referred to as CL,  can be made by mutating these two residues to cysteines. 
Disulfide bond is simulated by adding a FENE potential between  M16 and G121 (see Eq. \ref{eqn:potential}). The CL mutant allows us to assess the coupling between two distal residues, one in the Met20 loop and the other in the $\beta$F-$\beta$G loop, on the CS$\rightarrow$OS transition. Experiments by Benkovic and coworkers have established that mutation of G121 can affect the hydride transfer reaction \cite{Schiffer_Benkovic_Annu_Rev_06}. Finally, we also simulated the OS$\rightarrow$CS transition which occurs during the release of THF.

We used Brownian dynamics algorithm for which the characteristic time scale is $\tau_H = \tau_L \frac{\zeta \epsilon_h h}{k_B T}$. A typical value for $\tau_L$ for proteins is 3$ps$ \cite{Veitshans_FD_97}. The simulations were performed with $\zeta = 100 \tau_L^{-1}$ which corresponds to water viscosity.   The typical value of the integration step size is $h = 0.16 \tau_L$.  During the transition from one allosteric state to another, we reduced $h$ tenfold to maintain numerical stability.   We first equilibrate DHFR in the CS  at T=300K for 80 $\mu$s. Subsequently, the forces are computed using the OS state Hamiltonian.  The equations of motion are integrated for long times (typically exceeding 700 $\mu$s) so that the transition to the OS state is complete. A similar procedure is used for the reverse reaction, and the cysteine crosslink mutants.   We generated 50 trajectories for each conformational transition so that the reported results for the kinetics are statistically significant.

 {\bf Acknowledgment:} We are grateful to Prof. Sharon Hammes-Schiffer whose seminar at the University of Maryland inspired this work. We appreciate the interest shown by Prof. Arieh Warshel in this work and for his comments on an earlier version of the paper.We are also grateful to Gordon Hammes for his encouragement and useful discussions.  This work supported part by a grant from the National Science Foundation through grant number CHE05-14056.

\newpage

\begin{thebibliography}{10}
\bibliographystyle{unsrt}

\bibitem{Schnell_Annu_Rev_04}
Schnell, J.~R., Dyson, H.~J.  \& Wright, P.~E. (2004{\em{}}).
\newblock Structure, dynamics, and catalytic function of dihydrofolate
  reductase.
\newblock {\em Annu. Rev. Biophys. Biomol. Struct.} {\bf 33}, 119--140.

\bibitem{Boehr_Science_06}
Boehr, D.~D., McElheny, D., Dyson, H.~J.  \& Wright, P.~E. (2006{\em{}}).
\newblock The dynamic energy landscape of dihydrofolate reductase catalysis.
\newblock {\em Science, } {\bf 313} (5793), 1638--1642.

\bibitem{Hammes_Nature_64}
Hammes, G.~G. (1964{\em{}}).
\newblock Mechanism of enzyme catalysis.
\newblock {\em Nature, } {\bf 204}, 342--343.

\bibitem{Schiffer_Benkovic_Annu_Rev_06}
Hammes-Schiffer, S. \& Benkovic, S.~J. (2006{\em{}}).
\newblock Relating protein motion to catalysis.
\newblock {\em Annu. Rev. Biochem.} {\bf 75}, 519--541.

\bibitem{Hyeon_Biochem_05}
Thirumalai, D. \& Hyeon, C. (2005{\em{}}).
\newblock {RNA and protein folding: common themes and variations}.
\newblock {\em Biochemistry, } {\bf 44} (13), 4957--4970.

\bibitem{Schiffer_PNAS_02}
Agarwal, P.~K., Billeter, S.~R., Rajagopalan, P. T.~R., Benkovic, S.~J.  \&
  Hammes-Schiffer, S. (2002{\em{}}).
\newblock {Network of coupled promoting motions in enzyme catalysis}.
\newblock {\em Proc. Natl. Acad. Sci.} {\bf 99} (5), 2794--2799.

\bibitem{Schiffer_JPCB_02}
Agarwal, P.~K., Billeter, S.~R.  \& Hammes-Schiffer, S. (2002{\em{}}).
\newblock Nuclear quantum effects and enzyme dynamics in dihydrofolate
  reductase catalysis.
\newblock {\em J. Phys. Chem. B, } {\bf 106}, 3283--3293.

\bibitem{Schiffer_JPCB_04}
Wong, K., Watney, J.  \& Hammes-Schiffer, S. (2004{\em{}}).
\newblock Analysis of electrostatics and correlated motions for hydride
  transfer in dihydrofolate reductase.
\newblock {\em J. Phys. Chem. B, } {\bf 108} (32), 12231--12241.

\bibitem{Schiffer_PNAS_05}
Wong, K.~F., Selzer, T., Benkovic, S.~J.  \& Hammes-Schiffer, S. (2005{\em{}}).
\newblock Impact of distal mutations on the network of coupled motions
  correlated to hydride transfer in dihydrofolate reductase.
\newblock {\em Proc. Natl. Acad. Sci.} {\bf 102}, 6807--6812.

\bibitem{Benkovic_Biochem_02}
Rajagopalan, P. T.~R., Lutz, S.  \& Benkovic, S.~J. (2002{\em{}}).
\newblock Coupling interactions of distal residues enhance dihydrofolate
  reductase catalysis: mutational effects on hydride transfer rates.
\newblock {\em Biochemistry, } {\bf 41} (42), 12618--12628.

\bibitem{Cannon_NSB_96}
Cannon, W.~R., Singleton, S.~F.  \& Benkovic, S.~J. (1996{\em{}}).
\newblock A perspective on biological catalysis.
\newblock {\em Nat. Struct. Mol. Biol.} {\bf 3}, 821--833.

\bibitem{Brooks_JACS_00}
Radkiewicz, J.~L. \& C.~L.~Brooks, I. (2000{\em{}}).
\newblock Protein dynamics in enzymatic catalysis: exploration of dihydrofolate
  reductase.
\newblock {\em J. Am. Chem. Soc.} {\bf 122}, 225--231.

\bibitem{Brooks_PNAS_03}
{Rod T. H., Radkiewicz J. L.\& Brooks, C. L. III} (2003{\em{}}).
\newblock Correlated motion and the effect of distal mutations in dihydrofolate
  reductase.
\newblock {\em Proc. Natl. Acad. Sci.} {\bf 100}, 6980--6985.

\bibitem{Brooks_JPC_03}
{Thorpe, I. F. and Brooks, C. L. III} (2003{\em{}}).
\newblock Barriers to hydride transfer in wild type and mutant dihydrofolate
  reductase from e. coli.
\newblock {\em J. Phys. Chem. B, } {\bf 107}, 14042--14051.

\bibitem{Brooks_Proteins_04}
{Thorpe I. F. \& Brooks, C. L. III} (2004{\em{}}).
\newblock The coupling of structural fluctuations to hydride transfer in
  dihydrofolate reductase.
\newblock {\em Proteins: Structure, Function, and Bioinformatics, } {\bf 57},
  444--457.

\bibitem{Gao_Biochem_03}
Garcia-Viloca, M., Truhlar, D.  \& Gao, J. (2003{\em{}}).
\newblock Reaction-path energetics and kinetics of the hydride transfer
  reaction catalyzed by dihydrofolate reductase.
\newblock {\em Biochemistry, } {\bf 42} (46), 13558--13575.

\bibitem{Warshel_ChemRev_06}
Olsson, M., Parson, W.  \& Warshel, A. (2006{\em{}}).
\newblock Dynamical contributions to enzyme catalysis: critical tests of a
  popular hypothesis.
\newblock {\em Chem. Rev.} {\bf 106} (5), 1737--1756.

\bibitem{Benkovic_Biochem_97}
Cameron, C.~E.,   \& Benkovic, S.~J. (1997{\em{}}).
\newblock Evidence for a functional role of the dynamics of glycine-121 of
  escherichia coli dihydrofolate reductase obtained from kinetic analysis of a
  site-directed mutant.
\newblock {\em Biochemistry, } {\bf 36} (50), 15792--15800.

\bibitem{Benkovic_PNAS_06}
Wang, L., Goodey, N.~M., Benkovic, S.~J.  \& Kohen, A. (2006{\em{}}).
\newblock Coordinated effects of distal mutations on environmentally coupled
  tunneling in dihydrofolate reductase.
\newblock {\em Proc. Natl. Acad. Sci.} {\bf 103}, 15753--15758.

\bibitem{Feeney_Biochem_99}
Polshakov, V.~I., Birdsall, B.  \& Feeney, J. (1999{\em{}}).
\newblock Characterization of rates of ring-flipping in trimethoprim in its
  ternary complexes with lactobacillus casei dihydrofolate reductase and
  coenzyme analogues.
\newblock {\em Biochemistry, } {\bf 38} (48), 15962--15969.

\bibitem{Wright_Biochem_01}
Osborne, M.~J., Schnell, J., Benkovic, S.~J., Dyson, H.~J.  \& Wright, P.~E.
  (2001{\em{}}).
\newblock Backbone dynamics in dihydrofolate reductase complexes: role of loop
  flexibility in the catalytic mechanism.
\newblock {\em Biochemistry, } {\bf 40} (33), 9846--9859.

\bibitem{Wright_PNAS_05}
McElheny, D., Schnell, J.~R., Lansing, J.~C., Dyson, H.~J.  \& Wright, P.~E.
  (2005{\em{}}).
\newblock {Defining the role of active-site loop fluctuations in dihydrofolate
  reductase catalysis}.
\newblock {\em Proc. Natl. Acad. Sci.} {\bf 102} (14), 5032--5037.

\bibitem{Warshel_JPC_91}
Hwang, J.~K., Chu, Z.~T., Yadav, A.  \& Warshel, A. (1991{\em{}}).
\newblock Simulations of quantum mechanical corrections for rate constants of
  hydride-transfer reactions in enzymes and solutions.
\newblock {\em Journal of Physical Chemistry, } {\bf 95} (22), 8445--8448.

\bibitem{Warshel_QRB_01}
Warshel, A. \& Parson, W. (2001{\em{}}).
\newblock Dynamics of biochemical and biophysical reactions: insight from
  computer simulations.
\newblock {\em Quart. Rev. Biophys.} {\bf 34}, 563--679.

\bibitem{Warshel_JPC_01}
Villa, J. \& Warshel, A. (2001{\em{}}).
\newblock Energetics and dynamics of enzymatic reactions.
\newblock {\em Journal of Physical Chemistry B, } {\bf 105} (33), 7887--7907.

\bibitem{Warshel_Biochem_07}
Liu, H. \& Warshel, A. (2007{\em{}}).
\newblock The catalytic effect of dihydrofolate reductase and its mutants is
  determined by reorganization energies.
\newblock {\em Biochemistry, } {\bf 46}.

\bibitem{Benkovic_ChemBiol_98}
Miller, G.~P. \& Benkovic, S.~J. (1998{\em{}}).
\newblock Stretching exercises -- flexibility in dihydrofolate reductase
  catalysis.
\newblock {\em Chemistry $\&$ Biology, } {\bf 5}, R105--R113.

\bibitem{Berg_Biochem}
Berg, J., Tymoczko, J.~T.  \& Stryer, L. (2002{\em{}}).
\newblock {\em Biochemistry}.
\newblock Fifth edition, New York: W. H. Freeman and Co.

\bibitem{Matthews_Science_77}
Matthews, D., Alden, R., Bolin, J., Freer, S., Hamlin, R., Xuong, N., Kraut,
  J., Poe, M., Williams, M.  \& Hoogsteen, K. (1977{\em{}}).
\newblock {Dihydrofolate reductase: x-ray structure of the binary complex with
  methotrexate}.
\newblock {\em Science, } {\bf 197} (4302), 452--455.

\bibitem{Kraut_Biochem_97}
Sawaya, M. \& Kraut, J. (1997{\em{}}).
\newblock Loop and subdomain movements in the mechanism of escherichia coli
  dihydrofolate reductase: crystallographic evidence.
\newblock {\em Biochemistry, } {\bf 36}, 586--603.

\bibitem{Benkovic_Wright_Biochem_04}
Venkitakrishnan, R.~P., Zaborowski, E., McElheny, D., Benkovic, S.~J., Dyson,
  H.~J.  \& Wright, P.~E. (2004{\em{}}).
\newblock Conformational changes in the active site loops of dihydrofolate
  reductase during the catalytic cycle.
\newblock {\em Biochemistry, } {\bf 43} (51), 16046--16055.

\bibitem{RR_Science_99}
Lockless, S. \& Ranganathan, R. (1999{\em{}}).
\newblock Evolutionarily conserved pathways of energetic connectivity in
  protein families.
\newblock {\em Science, } {\bf 286}, 295--299.

\bibitem{Dima_ProtSci_06}
Dima, R.~I. \& Thirumalai, D. (2006{\em{}}).
\newblock Determination of network of residues that regulate allostery in
  protein families using sequence analysis.
\newblock {\em Protein Sci.} {\bf 15}, 258--268.

\bibitem{RR_NSB_03}
Suel, G., Lockless, S., Wall, M.  \& Ranganathan, R. (2003{\em{}}).
\newblock Evolutionarily conserved networks of residues mediate allosteric
  communication in proteins.
\newblock {\em Nat. Strul. Biol.} {\bf 10}, 59--68.

\bibitem{RR_PNAS_03}
Hatley, M., Lockless, S., Gibson, S., Gilman, A.  \& Ranganathan, R.
  (2003{\em{}}).
\newblock Allosteric determinants in guanine nucleotide-binding proteins.
\newblock {\em Proc. Natl. Acad. Sci.} {\bf 100}, 14445--14450.

\bibitem{RR_PNAS_04}
Jain, R. \& Ranganathan, R. (2002{\em{}}).
\newblock Local complexity of amino acid interactions in a protein core.
\newblock {\em Proc. Natl. Acad. Sci.} {\bf 101}, 111--116.

\bibitem{RR_cell_04}
Shulman, A., Larson, C., Mangelsdorf, D.  \& Ranganathan, R. (2004{\em{}}).
\newblock Structural determinants of allosteric ligand activation in rxr
  heterodimers.
\newblock {\em Cell, } {\bf 116}, 417--429.

\bibitem{Hyeon_BiophyJ_07}
Hyeon, C. \& Thirumalai, D. (2007{\em{}}).
\newblock Mechanical unfolding of {RNA}: from hairpins to structures with
  internal multiloops.
\newblock {\em Biophys. J.} {\bf 192}, 731--743.

\bibitem{Clore_Nature_06}
Tang, C., Iwahara, J.  \& Clore, G. (2006{\em{}}).
\newblock Visualization of transient encouter complexes in protein-protein
  association.
\newblock {\em Nature, } {\bf 444}, 383--386.

\bibitem{Pfam}
Bateman, A., Birney, E., Cerruti, L., Durbin, R., Etwiller, L., Eddy, S.,
  Griffiths-Jones, S., Howe, K., Marshall, M.  \& Sonnhammer, E. (2002{\em{}}).
\newblock The pfam protein families database.
\newblock {\em Nucleic Acids Res.} {\bf 30}, 276--280.

\bibitem{ClustalW}
Thompson, J., Higgins, D.  \& Gibson, T. (1994{\em{}}).
\newblock Clustal w: improving the sensitivity of progressive multiple sequence
  alignment through sequence weighting, position-specific gap penalties and
  weight matrix choice.
\newblock {\em Nucleic Acids Res.} {\bf 22}, 4673--4680.

\bibitem{Blatt_PRL_96}
Blatt, M., Wiseman, S.  \& Domany, E. (1996{\em{}}).
\newblock Superparamagnetic clustering of data.
\newblock {\em Phys. Rev. Lett.} {\bf 76} (18), 3251--3254.

\bibitem{Domany_PhysicaA_99}
Domany, E. (1999{\em{}}).
\newblock Superparamagnetic clustering of data - the definitive solution of an
  ill-posed problem.
\newblock {\em Physica A.} {\bf 263}, 158--169.

\bibitem{Stan_BiophyChem_03}
Stan, G., Thirumalai, D., Lorimer, G.~H.  \& Brooks, B.~R. (2003{\em{}}).
\newblock { Annealing. function of GroEL: structural and bioinformatic
  analysis.}
\newblock {\em Biophys. Chem.} {\bf 100}, 453--467.

\bibitem{Kraut_Science_86}
Howell, E., Villafranca, J., Warren, M., Oatley, S.  \& Kraut, J.
  (1986{\em{}}).
\newblock Functional role of aspartic acid-27 in dihydrofolate reductase
  revealed by mutagenesis.
\newblock {\em Science, } {\bf 231}, 1123--1128.

\bibitem{Benkovic_Philosophical_06}
Wang, L., Goodey, N.~M., Benkovic, S.~J.  \& Kohen, A. (2006{\em{}}).
\newblock The role of enzyme dynamics and tunnelling in catalysing hydride
  transfer: studies of distal mutants of dihydrofolate reductase.
\newblock {\em Phil. Trans. R. Soc. B, } {\bf 361}, 1307--1315.

\bibitem{Schiffer_JPCB_06_Freezing_Motion}
Sergi, A., Watney, J., Wong, K.  \& Hammes-Schiffer, S. (2006{\em{}}).
\newblock Freezing a single distal motion in dihydrofolate reductase.
\newblock {\em J. Phys. Chem. B, } {\bf 110} (5).

\bibitem{Tsai_ProtSci_99}
Tsai, C., Kumar, S., Ma, B.  \& Nussinov, R. (1999{\em{}}).
\newblock {Folding funnels, binding funnels, and protein function}.
\newblock {\em Protein Sci, } {\bf 8} (6), 1181--1190.

\bibitem{Koshland_PNAS_58}
Koshland, D.~E. (1958{\em{}}).
\newblock {Application of a Theory of Enzyme Specificity to Protein Synthesis}.
\newblock {\em Proc. Natl. Acad. Sci.} {\bf 44} (2), 98--104.

\bibitem{Changeux_Science_05}
Changeux, J. \& Edlestein, S. (2005{\em{}}).
\newblock Allosteric mechanism of signal transduction.
\newblock {\em Science, } {\bf 308}, 1424--1428.

\bibitem{Warshel_JPCB_07}
Liu, H. \& Warshel, A. (2007{\em{}}).
\newblock Origin of the temperature dependence of isotope effects in enzymatic
  reactions: the case of dihydrofolate reductase.
\newblock {\em Journal of Physical Chemistry B, } {\bf 111} (27), 7852--7861.

\bibitem{Hyeon_Struct_06}
Hyeon, C., Dima, R.~I.  \& Thirumalai, D. (2006{\em{}}).
\newblock Pathways and kinetic barriers in mechanical unfolding and refolding
  of rna and proteins.
\newblock {\em Structure, } {\bf 14}, 1633--1645.

\bibitem{Hyeon_PNAS_06}
{Hyeon}, C., {Lorimer}, G.~H.  \& {Thirumalai}, D. (2006{\em{}}).
\newblock {Dynamics of allosteric transitions in GroEL}.
\newblock {\em Proc. Natl. Acad. Sci.} {\bf 103}, 18939--18944.

\bibitem{Domany_PNAS_00}
Getz, G., Levine, E.  \& Domany, E. (2000{\em{}}).
\newblock Coupled two-way clustering analysis of gene microarray data.
\newblock {\em Proc. Natl. Acad. Sci.} {\bf 97}, 12079--12084.

\bibitem{Hyeon_PNAS_07}
Hyeon, C. \& Onuchic, J.~N. (2007{\em{}}).
\newblock Internal strain regulates the nucleotide binding site of the kinesin
  leading head.
\newblock {\em Proc. Natl. Acad. Sci.} {\bf 104}, 2175.

\bibitem{Gardinerbook}
Gardiner, C.~W. (1996{\em{}}).
\newblock {\em Handbook of stochastic methods for physics, chemistry, and the
  natural sciences}.
\newblock Second edition, Springer.

\bibitem{Veitshans_FD_97}
Veitshans, T., Klimov, D.  \& Thirumalai, D. (1997{\em{}}).
\newblock Protein folding kinetics: timescales, pathways and energy landscapes
  in terms of sequence-dependent properties.
\newblock {\em Fold Des.} {\bf 2}, 1--22.

\end{thebibliography}

\newpage
\section*{\bf Figure Captions}

\begin{figure}[ht]
\includegraphics[width=6.00in]{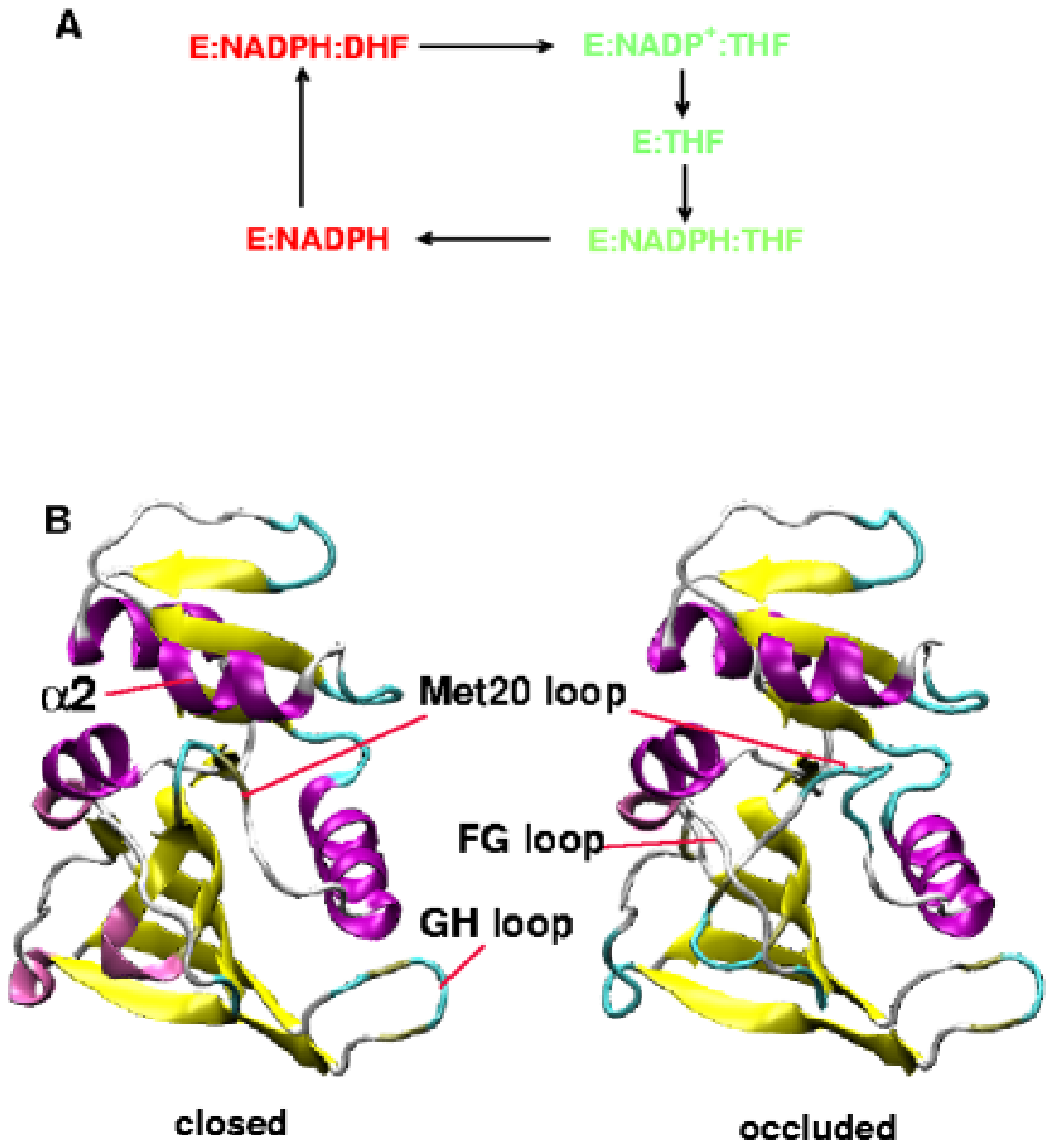}
\caption{\label{fig:catalysis_cycle}}
\end{figure}

{{\bf Figure} \ref{fig:catalysis_cycle} :}
Catalytic cycle and structures of closed and occluded states. (A) Scheme of catalytic cycle of DHFR that shows the two key conformations adopted by the enzyme. The Met20 loop closes the active loop in the E:NADPH:DHF complex, while it is occluded in the E:NADP$^{+}$:THF complex. (B) Structure of the closed state (CS) (PDB code 1RX2) is on the left and the occluded state (OS) (PDB code 1RX7) is on the right. For clarity, we have explicitly labeled the structural elements that facilitate the allosteric transitions. The major changes are localized in the Met20 loop. 

\begin{figure}[ht]
\includegraphics[height=9in]{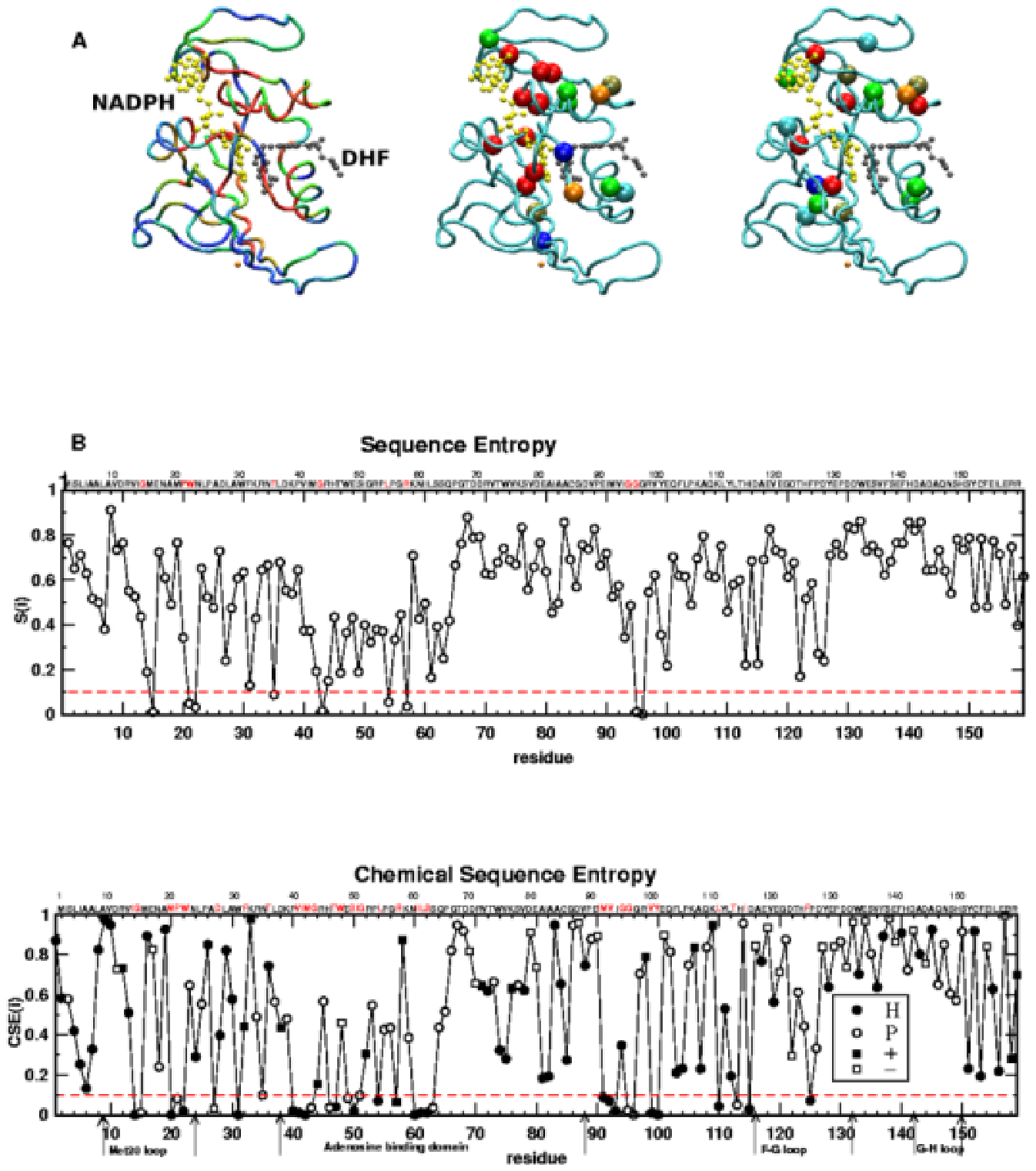}
\caption{\label{fig:conservation}}
\end{figure}

{{\bf Figure} \ref{fig:conservation} :}
Sequence conservation with respect to the structure of the CS. (A) In the CS structure on the left we have color coded the backbone to reflect the extent of sequence conservation. Red color represents strong conservation ($S(i)<0.1$) and non-conserved residues are in blue. The residues that are clustered in the position in the SCA (V13, I14, N18, M20, D27, L28, K38, V40, I41, M42, W47, I50, G51, N59, I60, L62, Q65, I94, V99, Y111 and T113) are shown in the middle structure. The colors of the residues indicate the values of the $S_{CSE}(i)$ with red representing strong conservation of the chemical identity. The right structure shows the network of residues that appear as perturbation in the SCA (D11, R12, V13, D27, A29, K38, V40, W47, I50, G51, P53, L62, V72, S77, A84, G97, V99, Q108, and Y111). In the three structures the cofactor (NADPH and DHF) are shown using all-atom representation. (B) The top panel shows position dependent sequence entropy ($S(i)$) obtained by aligning the {\it E. coli} DHFR against the rest of the 426 sequences. Strongly conservation ($S(i)<0.1$) is observed only for a small fraction of residues. The chemical sequence entropy, $S_{CSE}(i)$ in the bottom panel, shows that for a substantial fraction of residues only the chemical identity changes. These residues are dispersed throughout the structure. 

\begin{figure}[ht]
\includegraphics[width=7.0in]{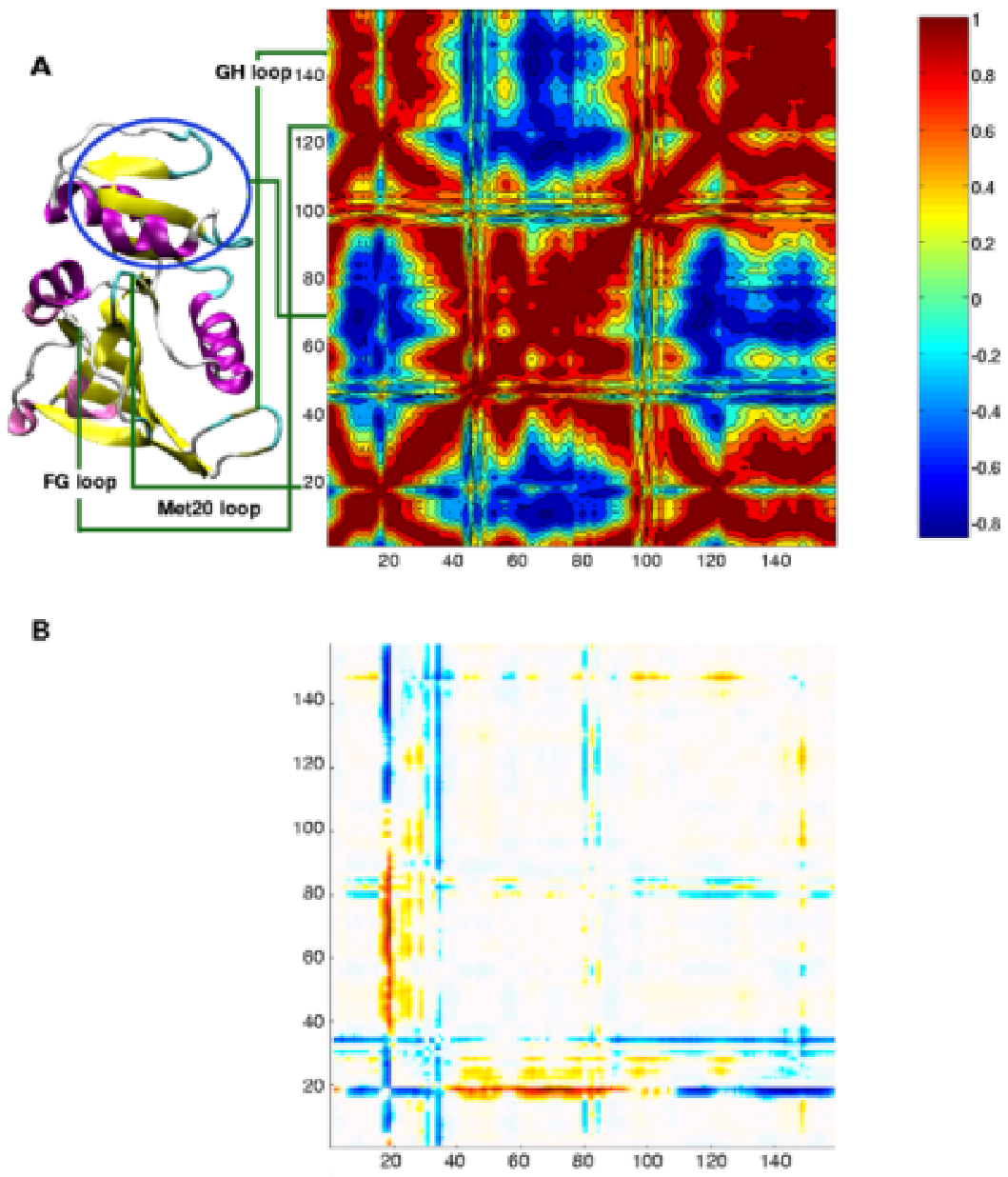}
\caption{\label{fig:correlation}}
\end{figure}

{{\bf Figure} \ref{fig:correlation} :}
Correlated and anti-correlated motions in DHFR. Covariance matrix of equilibrium fluctuations of the unit vectors constructed from the coordinates of the $C_\alpha$ atoms (Eq. \ref{eq:correlation}) of the wild type DHFR in the CS. The residues associated with the structural elements are shown on the left. The scale on the right measures the extent of correlation. In the bottom panel we show $|C_{ij}^{CS}-C_{ij}^{OS}|$ using the same scale. For clarity, we only highlight those regions that are different in the two allosteric states. The simplicity of the SOP model has allowed us to probe the equilibrium fluctuations on about 0.1 msec.

\begin{figure}[ht]
\includegraphics[width=6in]{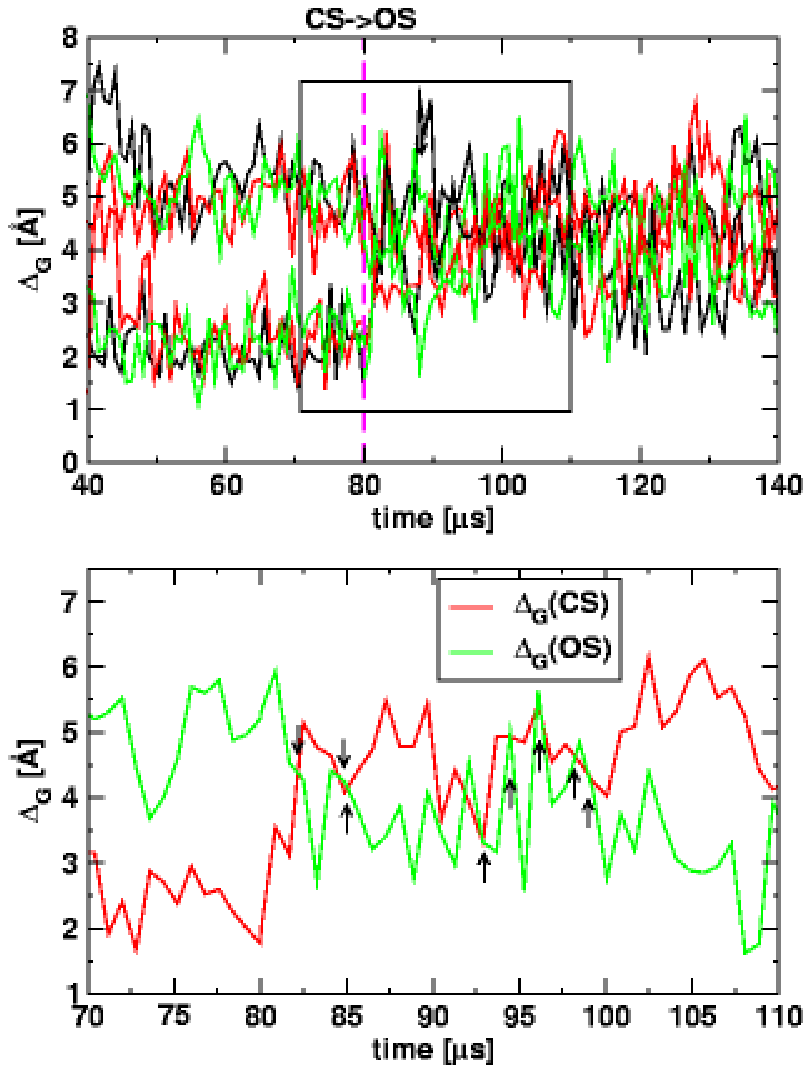}
\caption{\label{fig:multiple_crossing}}
\end{figure}

{{\bf Figure} \ref{fig:multiple_crossing} :}
RMSD as a function of time during the CS$\rightarrow$OS transition. Time dependent changes in the global RMSD ($\Delta_G(t)$) for a few representative trajectories as a function of $t$ are given in the top panel. The dynamical changes in $\Delta_G(t)$ for one of the trajectories for the time interval enclosed by the box are shown below. The arrows show that frequent recrossings between CS and OS states occur prior to the completion of CS$\rightarrow$OS transition.

\begin{figure}[ht]
\includegraphics[width=7in]{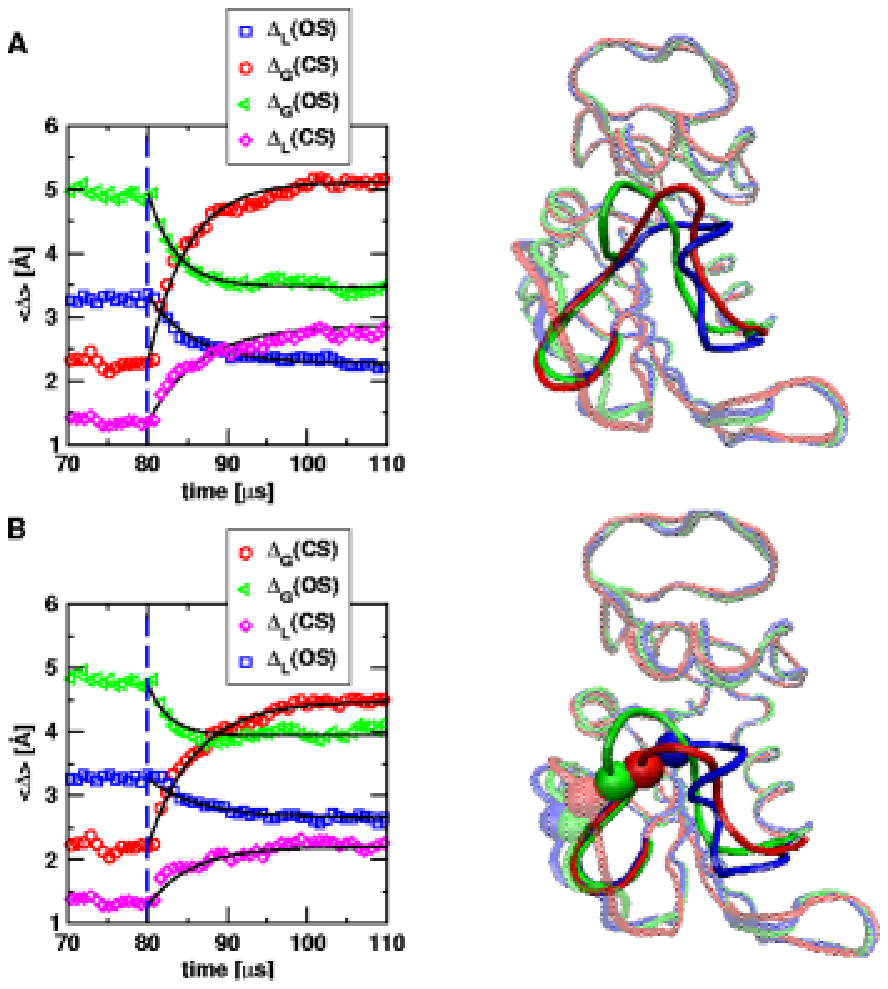}
\caption{\label{fig:forward_transition}}
\end{figure}

{{\bf Figure} \ref{fig:forward_transition} :}
Kinetics of allosteric transitions probed using RMSD. Time dependent changes in the global, $<\Delta_G(t)>$, and the local, $<\Delta_L(t)>$, RMSD of the Met20 loop averaged over 50 trajectories. The meaning of the symbols are given in the insets. The RMSDs are measured with respect to the starting and ending states. For example, $<\Delta_G(CS)>$ means that the global RMSD is computed with respect to the CS. The changes in $<\Delta_G(t)>$ and $<\Delta_L(t)>$ for the WT CS$\rightarrow$OS are shown in (A), and (B) shows the results for the CL. The structures on the right in both (A) and (B) represent superposition of the CS, OS and the average transition state (TS) conformations. Conformation of the Met20 loop in the CS (green), OS (blue), and TS (red) are highlighted. The cross link between 16 and 121 is explicitly shown in (B).  

\begin{figure}[ht]
\includegraphics[width=7in]{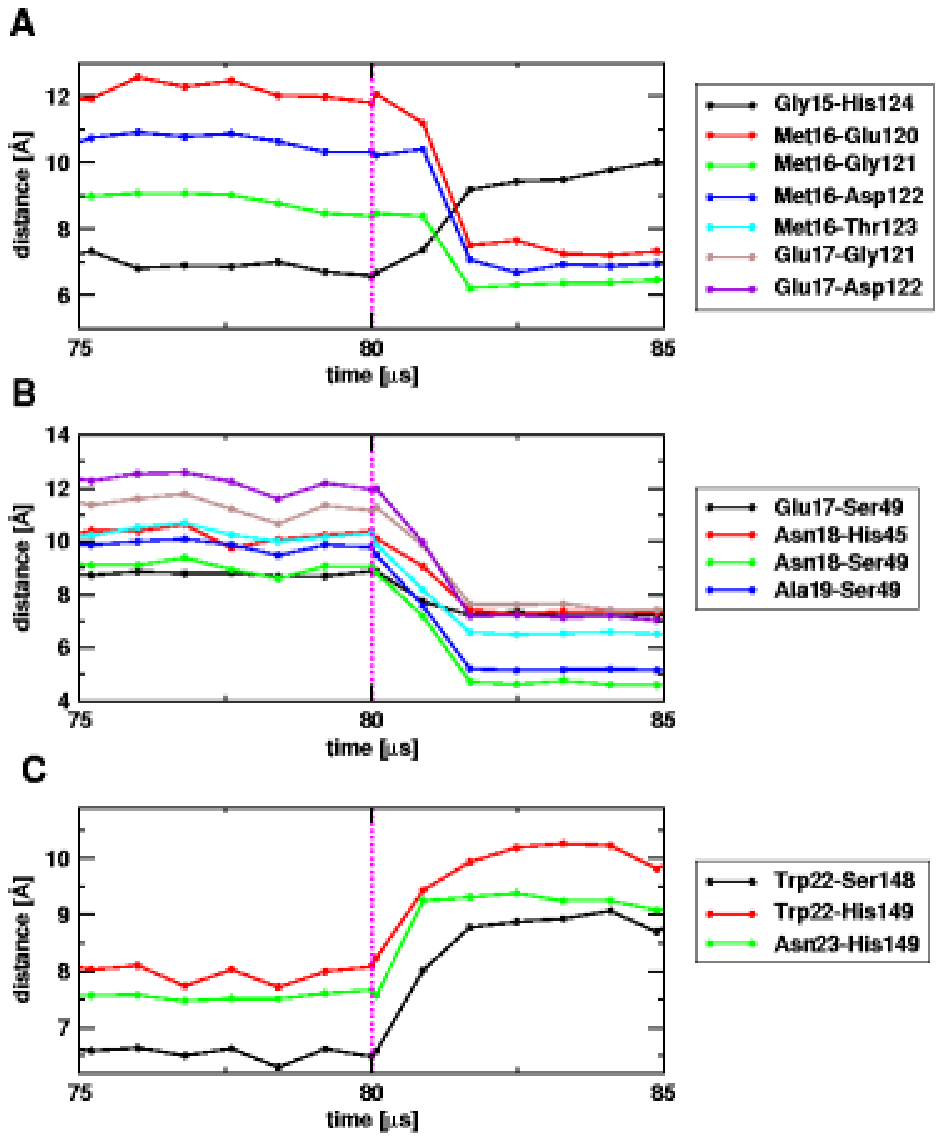}
\caption{\label{fig:contacts}}
\end{figure}

{{\bf Figure} \ref{fig:contacts} :}
Kinetics of rupture of individual contacts between residues in the various substructures of DHFR during the CS$\rightarrow$OS transition. (A) This panel shows the changes in the distance between the residues in the Met20 loop and $\alpha$2 that rupture and form. (B) Same as (A) except the residues represent interactions between the Met20 loop and $\beta$F-$\beta$G loop. (C) shows the kinetics describing the formation of interactions between the Met20 loop and $\beta$G-$\beta$H loop. In all cases the identifies of the residues are shown on the right. 

\begin{figure}[ht]
\includegraphics[width=7in]{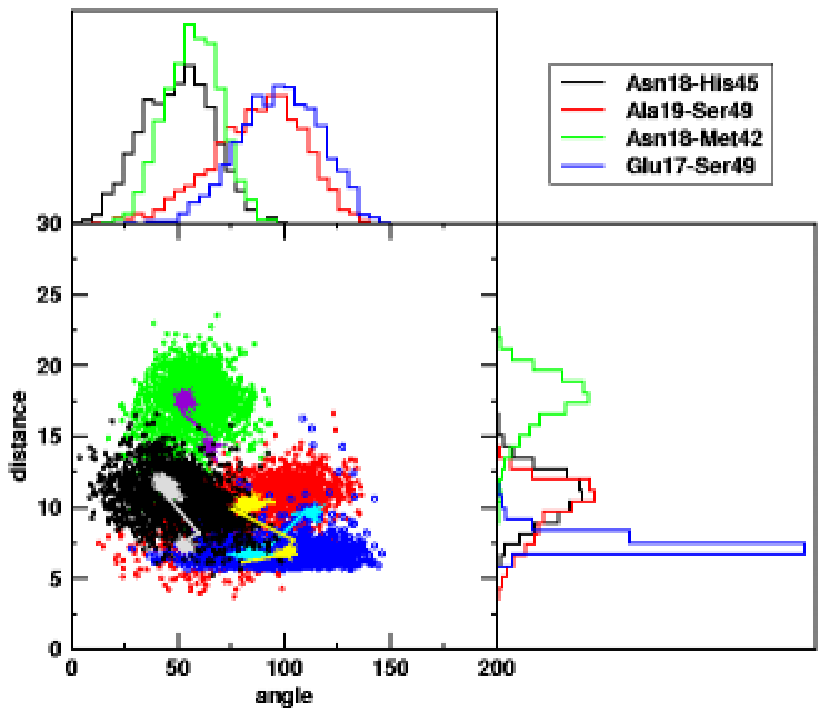}
\caption{\label{fig:R_alpha}}
\end{figure}

{{\bf Figure} \ref{fig:R_alpha} :}
Illustration of the sliding of the Met20 loop along $\alpha$2. Two dimensional projection of the distance ($R_i$) and angle ($\alpha_i$) that are kinetically sampled in the 50 trajectories during the CS$\rightarrow$OS transition. The angles $\alpha_i$ are defined in the text. The monitored residues are identified on the upper right corner. The colors grey, yellow, purple and cyan represent ensemble averages. The monotonic decrease in all the angles shows sliding of the Met20 loop along $\alpha$2. Histograms of the distance and the angles for the residue pairs are displayed on the right and on top respectively. 

\begin{figure}[ht]
\includegraphics[width=7in]{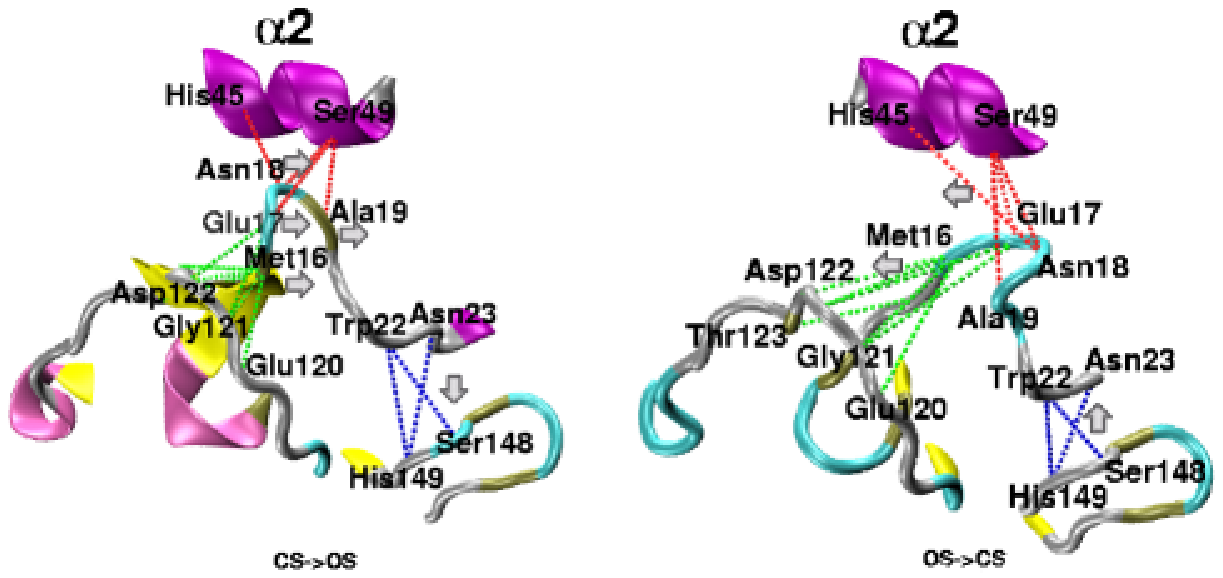}
\caption{\label{fig:Motion}}
\end{figure}

\begin{figure}[ht]
\includegraphics[width=8in]{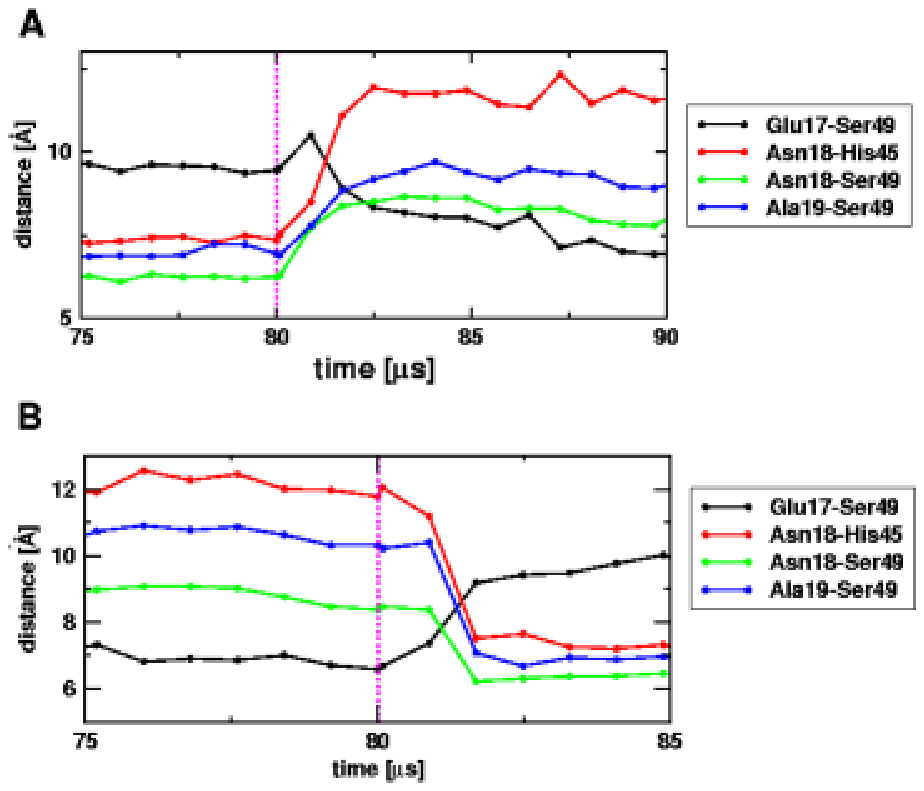}
\caption{\label{fig:contacts_CL}}
\end{figure}

{{\bf Figure} \ref{fig:Motion} :}
Structural representation of the coordinated changes in the distances of residues that accompany the sliding motion in the CS$\rightarrow$OS (left side) and OS$\rightarrow$CS (right side) transition. The arrows indicate the direction in which the structural changes occur. The displayed structural changes were inferred from the kinetics shown in Figs. \ref{fig:contacts}, \ref{fig:contacts_CL}, and \ref{fig:R_alpha}. In the forward transition the Met20 loop is pulled by the $\beta$G-$\beta$H loop which results in it being pushed away from $\beta$F-$\beta$G loop. The push-pull process results in the sliding of the Met20 loop. The mechanism is approximately reserved in the OS$\rightarrow$CS transition.

\begin{figure}[ht]
\includegraphics[width=7in]{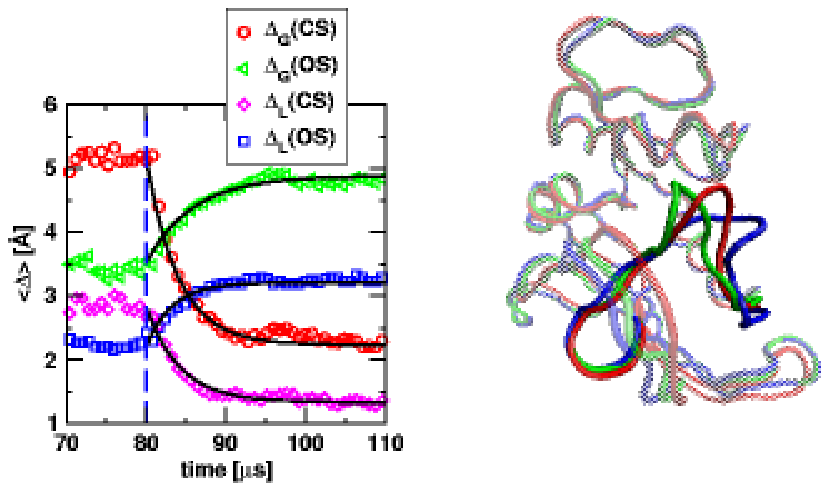}
\caption{\label{fig:reverse_transition}}
\end{figure}

{{\bf Figure} \ref{fig:reverse_transition} :}
Changes in $\Delta_G(t)$ and $\Delta_L(t)$ for the OS$\rightarrow$CS transitions. Same as Fig. \ref{fig:forward_transition} except that the transition is from OS$\rightarrow$CS. The conformation of the Met20 loop in the TS (red) is different from that in Fig. \ref{fig:forward_transition}.

{{\bf Figure} \ref{fig:contacts_CL} :}
Dissection of the local changes in the kinetics in CL (CS$\rightarrow$OS) and WT (OS$\rightarrow$CS). (A) Time-dependent changes in the distances between select residues from the Met20 loop and $\alpha$2 for the CS$\rightarrow$OS transition in the CL. The transition is from CS$\rightarrow$OS. (B) Same as (A) except these represent changes that occur during the OS$\rightarrow$CS transition for the WT.

\begin{figure}[ht]
\includegraphics[width=8in]{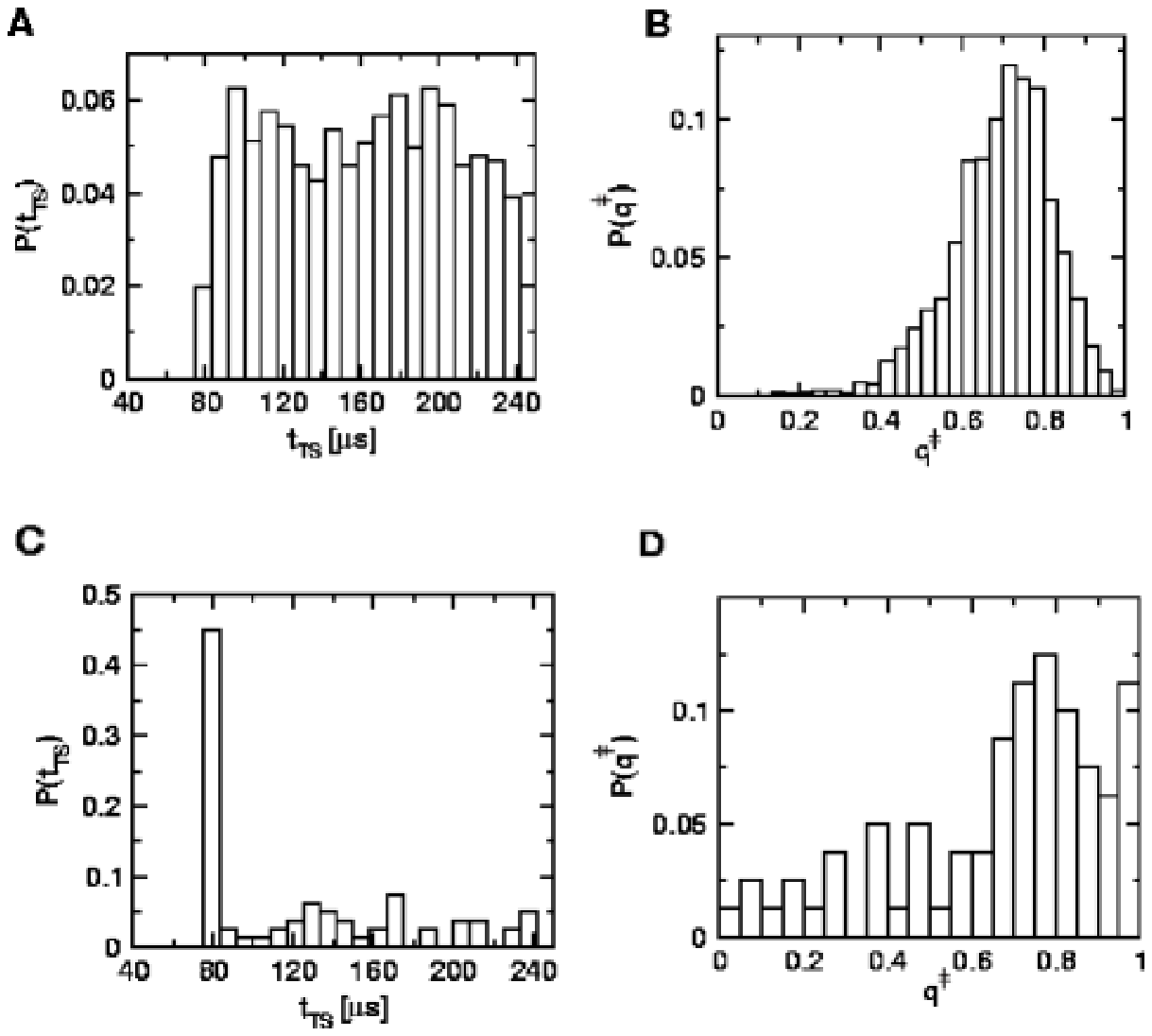}
\caption{\label{fig:TS_distribution}}
\end{figure}

{{\bf Figure} \ref{fig:TS_distribution} :}
Characteristics of the transition state ensemble (TSE). (A) and (C) show the distribution of transition times $P(t_{TS})$ for the forward and reverse transition, respectively. (B) and (D) represent the distributions $P(q^\ddagger)$ of $q^{\ddagger}$ (see Eq. \ref{eq:qdagger}) for the CS$\rightarrow$OS and OS$\rightarrow$CS transitions respectively. In both cases the TSE is broad. However, the width of the TSE (inferred from $P(q^{\ddagger})$) in the reverse direction is larger. The fluctuation $\frac{<q^{\ddagger 2}>-<q^\ddagger>^2}{<q^\ddagger>^2}$ is 0.2 for CS$\rightarrow$OS and 0.6 for OS$\rightarrow$CS.

\begin{figure}[ht]
\includegraphics[width=6in]{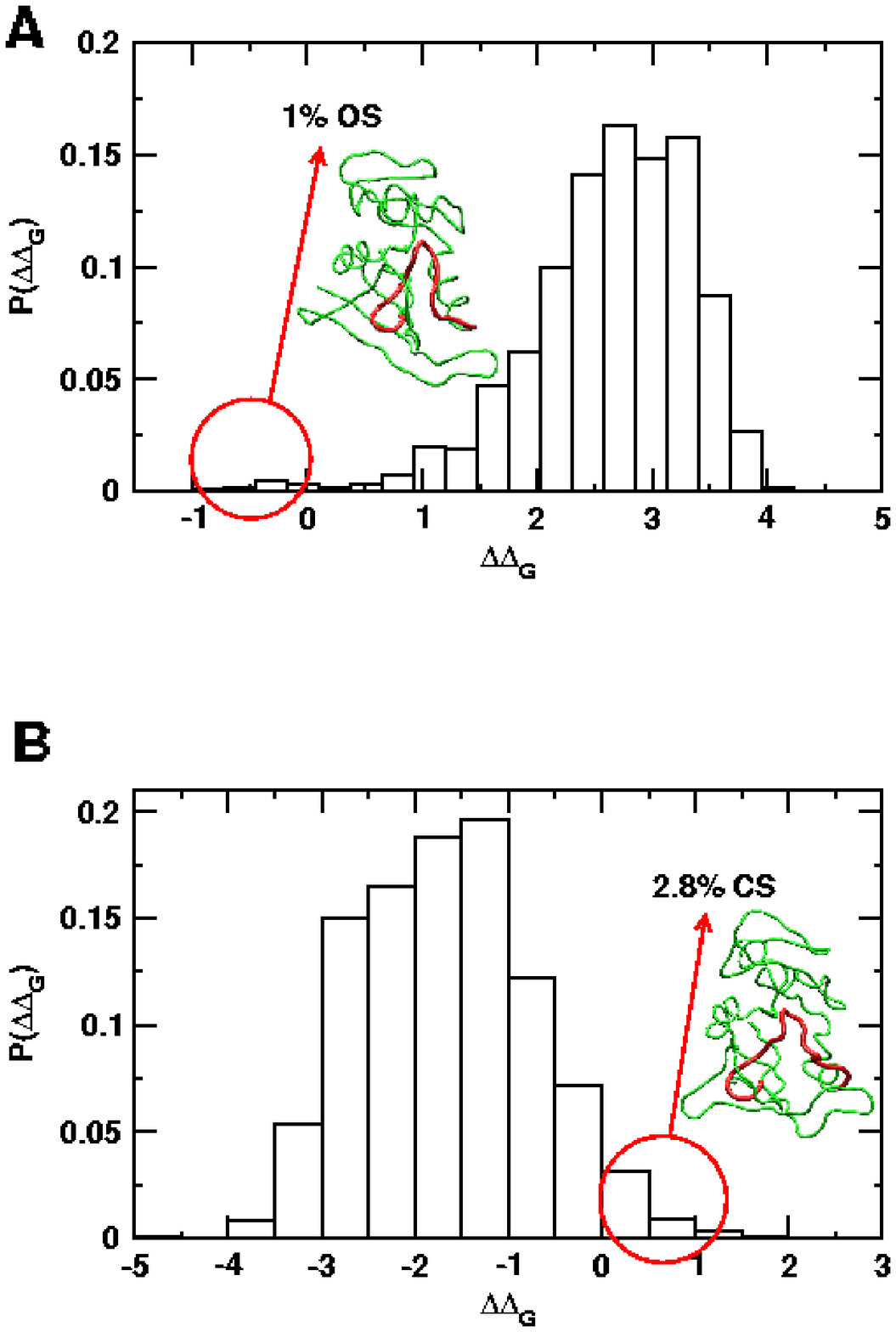}
\caption{\label{fig:pre_eq}}
\end{figure}

{{\bf Figure} \ref{fig:pre_eq} :}
Sampling of OS (CS) in CS (OS) state. (A) Distribution of $P(\Delta\Delta_G)$, calculated using an ensemble of equilibrated conformations in the CS, as a function of $\Delta\Delta_G$ (Eq. \ref{eq:ddg}). The negative regions represent sampling of conformations that resemble the OS. (B) Same as (A) except $P(\Delta\Delta_G)$ is obtained from an ensemble of equilibrated structures in OS state. Under equilibrium conditions a minor population $\sim (1-3) \%$ of the product-like structures are present. The displayed structure in (A) is OS-like while the one in (B) is CS-like.

\newpage

\newpage

\newpage

\newpage

\newpage

\newpage

\newpage

\newpage

\newpage

\newpage

\newpage

\newpage

\end{document}